\documentclass{aa}  

\DeclareUnicodeCharacter{FEFF}{}
\usepackage[]{natbib}
\usepackage[T1]{fontenc}
\usepackage{ae,aecompl}
\usepackage[dvipsnames]{xcolor}
\usepackage{psfrag,color}
\usepackage[]{inputenc}
\usepackage{graphicx}	
\usepackage{amsmath}	
\usepackage{pifont}
\usepackage{amssymb}	
\usepackage{comment}
\usepackage{longtable}
\usepackage{subfigure}
\usepackage{multirow}
\usepackage{adjustbox}
\usepackage{ulem}       
\usepackage{newtxmath} 
\usepackage{newtxtext}
\usepackage{pdflscape}
\usepackage{supertabular}
\usepackage{float}
\usepackage{rotating}
\usepackage{footnote}
\usepackage{times}
\usepackage{chemmacros}
\usepackage{xspace}

\usepackage[colorlinks=true, urlcolor=blue, linkcolor=blue, citecolor=blue]{hyperref} 

\makeatletter
\renewcommand*\aa@pageof{, page \thepage{} of \pageref*{LastPage}}
\makeatother

\definecolor{lime}{HTML}{A6CE39}
\DeclareRobustCommand{\orcidicon}{
	\begin{tikzpicture}
	\draw[lime, fill=lime] (0,0) 
	circle [radius=0.16] 
	node[white] {{\fontfamily{qag}\selectfont \tiny ID}};
	\draw[white, fill=white] (-0.0625,0.095) 
	circle [radius=0.007];
	\end{tikzpicture}
	\hspace{-2mm}
}

\foreach \x in {A, ..., Z}{\expandafter\xdef\csname orcid\x\endcsname{\noexpand\href{https://orcid.org/\csname orcidauthor\x\endcsname}
			{\noexpand\orcidicon}}
}

\usepackage[]{ulem}  
\newcommand\redout{\bgroup\markoverwith
{\textcolor{red}{\rule[0.5ex]{2pt}{0.8pt}}}\ULon}

\newcommand{\hi}{H\,\textsc{i}}
\newcommand{\hii}{H\,\textsc{ii}}
\newcommand{\oii}{O\,\textsc{ii}}

\newcommand{\foii}{[O\,\textsc{ii}]}
\newcommand{\foiii}{[O\,\textsc{iii}]}
\newcommand{\fnii}{[N\,\textsc{ii}]}

\newcommand{\fcliii}{[Cl\,\textsc{iii}]}

\newcommand{\fsii}{[S\,\textsc{ii}]}
\newcommand{\fsiii}{[S\,\textsc{iii}]}

\newcommand{\fariii}{[Ar\,\textsc{iii}]}
\newcommand{\fariv}{[Ar\,\textsc{iv}]}
\newcommand{\ffeiii}{[Fe\,\textsc{iii}]}

\newcommand{\nel}{$n_{\rm e}$} 
\newcommand{\tel}{$T_{\rm e}$}

\begin{document}

   \title{The DESIRED electron temperature relations in star-forming regions of the local Universe}

   \subtitle{}

   \author{M. Orte-Garc\'{\i}a\inst{1,2}{\orcidA{}}
          \and
          C. Esteban\inst{1,2}{\orcidB{}}
          \and
          J. Garc\'{\i}a-Rojas\inst{1,2}{\orcidC{}} 
          \and
          J. E. M\'endez-Delgado\inst{3}{\orcidD{}}
          \and
          K. Z. Arellano-C\'ordova\inst{4}{\orcidG{}}
        \and
        A. Z. Lugo-Aranda\inst{5}{\orcidH{}}
        \and
        L. Toribio San Cipriano\inst{6}{\orcidJ{}}
        \and
          F. F. Rosales-Ortega\inst{7}{\orcidE{}}
        \and
        I. R. Martínez-Hernández\inst{8}{\orcidI{}}
        \and
        E. Reyes-Rodr\'{\i}guez\inst{1,2}{\orcidF{}}
 }   

   \institute{Instituto de Astrof\'isica de Canarias, E-38205 La Laguna, Tenerife, Spain \email{maialen.orte@iac.es} 
         \and
             Departamento de Astrof\'isica, Universidad de La Laguna, E-38206 La Laguna, Tenerife, Spain 
        \and
            Instituto de Astronom\'ia, Universidad Nacional Aut\'onoma de M\'exico, Apartado Postal 70-264, Coyoac\'an, 04510, Mexico City, Mexico.
             \and
             Institute for Astronomy, University of Edinburgh, Royal Observatory, Edinburgh, EH9 3HJ, United Kingdom
             \and
             Instituto de Astronomía, Universidad Nacional Autónoma de México, A.P. 106, Ensenada 22800, BC, México
             \and
            Centro de Investigaciones Energ\'eticas, Medioambientales y Tecnol\'ogicas (CIEMAT), Madrid, Spain
        \and
             Instituto Nacional de Astrof\'isica, \'Optica y Electr\'onica (INAOE-CONAHCyT), Luis E. Erro 1, 72840, Tonantzintla, Puebla, Mexico  
        \and
             Departamento de Astronom\'ia y Astrof\'isica, Facultad de Ciencias Espaciales, Universidad Nacional Aut\'onoma de Honduras, Bulevar Suyapa, Tegucigalpa, M.D.C, Honduras}

   \date{Received ; accepted }

\authorrunning {Orte-Garc\'{\i}a et al.}
\titlerunning {Temperature relations in star-forming regions}
   \date{\today}

 \abstract
   {Accurate determinations of chemical abundances in \hii\ regions and star-forming galaxies rely critically on robust measurements of the electron temperature (\tel) across different ionic species. However, when using a two- or three-zone scheme to approximate the ionisation structure, it is usually possible to measure only a single temperature diagnostic, which makes it necessary to rely on empirical or theoretical relations to infer the temperatures in the remaining zones.}
   {We present a homogeneous observational study of  electron temperature ({\tel}) relations between different ionic species: \tel(\fnii), \tel(\foii), \tel(\foiii), \tel(\fsii), \tel(\fsiii) and \tel(\fariii), using a sample of 699 spectra of Galactic and extragalactic {\hii} regions and star-forming galaxies (SFGs) of the local Universe.}
   {We use the DEep Spectra of Ionised REgions Database Extended project (DESIRED-E) that comprises more than 3000 spectra of {\hii} regions and SFGs with direct determinations of \tel. From this database we select those spectra where it is possible to determine {\tel} for at least two ionic species. We recalculate the electron density (\nel) and \tel\ using updated atomic data and a fully consistent methodology. The resulting \tel--\tel\ relations were examined using orthogonal distance regression fits, their total and intrinsic dispersions quantified, and the fitted slopes compared with results from previous works and predictions from photoionisation models.}
   {Relations involving low-ionisation \tel\ diagnostics show large intrinsic dispersions, particularly those based on \tel(\foii) and \tel(\fsii), likely due to their higher sensitivity to \nel\ inhomogeneities, possible recombination contributions, and observational uncertainties. Relations using \tel(\fnii) exhibit systematically lower dispersions, indicating that this diagnostic provides a more reliable estimate of the low-ionisation zone temperature when only higher-ionisation diagnostics are available, despite its observational difficulty at low metallicity.  Overall, the fitted slopes are broadly consistent with photoionisation model predictions, particularly for relations with low intrinsic dispersion, shuch as those involving \tel(\fnii) and \tel(\fsiii). This work provides the most comprehensive observational characterisation to date of \tel--\tel\ relations in local {\hii} regions. The empirical relations presented here offer a robust basis for estimating \tel, especially when only a single diagnostic is available.}
   {}

   \maketitle
%

\section{Introduction}
\label{sec:introduction}
Chemical abundance calculations in ionised nebulae are primarily limited by the precision with which electron temperature (\tel) can be inferred. \tel\ is usually determined from the ratio of auroral to nebular collisionally excited lines (CELs) of certain ions present in the spectra of ionised nebulae. This is because the excitation energies of the upper atomic levels that give rise to auroral lines are higher than those of nebular lines, so their intensity is much more dependent on the mean energy of the free electrons whose  collisions populate those upper levels and, therefore, on \tel. Optical auroral CELs -- such as \foiii~$\lambda$4363 , \fnii~$\lambda5755$, or \fsiii~$\lambda 6312$, among others -- are normally rather faint and difficult to measure in distant and/or low-\tel\ ionised nebulae. Using \tel\ measurements to determine ionic and, ultimately, total abundances is what we call the ``direct'' method \citep[e.g.][]{Dinerstein:90,Berg:15}. With the use of ever-larger ground- and space-based telescopes and ever-more sensitive spectrographs and detectors, the direct method can apply to an increasing number of objects, even extremely distant galaxies observed with the \textit{James Webb Space Telescope}, JWST \citep[e.g.][]{ArellanoCordova:22, ArellanoCordova:25, Schaerer:22, Curti:23,Trump:23}. \foiii~$\lambda$4363 is usually the brightest auroral line observed in the optical spectra of star-forming regions. This is because O is the most abundant heavy element in cosmic objects and O$^{++}$ is the most common species for the ionisation conditions typical in these types of objects, especially in low-metallicity ones.  

It is well known that assuming nebulae as isothermal systems is not a realistic approximation. The assumption of two different \tel\  values for the high and low ionisation zones has been a very common method for determining ion abundances since pioneering works such as, for example, \citet{Peimbert:69}. This is because, according to photoionisation models, in radiation-bounded ionised nebulae, and especially in high-metallicity ones, \tel\ grows outwards due to the combination of the hardening of the ionising spectrum with increasing photoionisation optical depth and the strong cooling produced by the fine structure lines of \foiii\ in the inner parts of the nebula, where O$^{++}$ dominates \citep[e.g][]{Stasinska:80}. In this two-zones approximation, the two most common \tel\ indicators used are \tel(\foiii) for the high-ionisation zone and \tel(\fnii) -- or \tel(\foii) --  for the low-ionisation zone. Some authors also propose the use of a three-zone scheme when \tel(\fsiii) is available, which is considered to represent an intermediate-ionisation zone \citep[e.g.][]{Berg:20}. The determination of \tel(\fsiii) requires observations in the redder part of the optical spectrum to measure the nebular \fsiii~$\lambda \lambda9069, 9531$ lines which, on the other hand, are greatly affected by telluric spectral features \citep{Noll:12}. 

Based on the discussion above, there is a fairly general consensus to use the two-zones approximation -- or even three-zones if possible, although much less frequently -- to determine abundances in star-forming regions. However, \foiii~$\lambda$4363 is usually the only auroral line that is measured in most star-forming regions, especially in extragalactic \hii\ regions and definitely in star-forming galaxies (SFGs). In these cases, the so-called \tel--\tel\ relations are used to estimate the \tel\ of the low-ionisation zone and thus be able to apply the two-zone approximation. These relations can be constructed from the results of photoionisation models, but also from observations of star-forming regions where we have determinations of the two \tel\ indicators involved in the particular \tel--\tel\ relation we intend to apply. There are also situations or groups of objects in which the \tel\ of the low-ionisation zone is the only one available and the \tel\ representative of the high-ionisation zone must be estimated using \tel--\tel\ relations. This is the case of small Galactic \hii\ regions ionised by early B- or late O-type stars \citep{Esteban:18,ArellanoCordova:21}, but also of many of the \hii\ regions in local spiral galaxies analysed in the CHAOS project \citep{Berg:15, Berg:20, Croxall:16, Rogers:21,Rogers:22}. Another situation where \tel--\tel\ relations must be applied to obtain the \tel\ representative of the high-ionisation zone is when the spectrograph does not cover the spectral region of the \foiii~$\lambda$4363 line, as it happens with the Multi Unit Spectroscopic Explorer (MUSE) at the 8m Very Large telescope (VLT), where observations usually only allow determining \tel(\fsiii), \tel(\fnii), or both \citep[e.g.][]{Groves:23,Brazzini:24,RickardsVaught:24}.

There are many works in the literature that address and provide \tel--\tel\ relations for star-forming regions, using both photoionisation models and observational data. \citet{Campbell:86} were the first to propose a relation  between \tel(\foii) and \tel(\foiii) based on the photoionisation models of \citet{Stasinska:82}, while \citet{Garnett:92} used his own models to propose linear \tel--\tel\ relations for different \tel\ diagnostics. Subsequent works such as those by, for example, \citet{Pagel:92}, \citet{Thuan:95}, \citet{Izotov:97b}, \citet{Oey:00},  \citet{Deharveng:00}, or \citet{MendezDelgado:23b}  present relations for different \tel\ pairs using each of them results from different photoionisation models, such as those by \citet{Stasinska:90}, \citet{Stasinska:97},  \citet{sutherland:93}, or \citet{ValeAsari:16}. Studies examining \tel--\tel\ relations from an observational perspective have multiplied in recent decades due to the increased sensitivity of instrumentation and the aperture of available telescopes, which are able to detect faint auroral lines in an ever-increasing number of objects. There are a large number of studies that use samples from extragalactic \hii\ regions in nearby spiral galaxies and/or dwarf galaxies of the Local Group \citep[e.g.][]{Vermeij:02, Kennicutt:03, Berg:15, Croxall:16, Berg:20, Rogers:21, Rogers:22, Zurita:21, Scholte:26} or samples including both, \hii\ regions and SFGs \citep[e.g.][]{Pilyugin:06, Pilyugin:09, Hagele:06, Esteban:09, Yates:20, MendezDelgado:23b, Cataldi:25} to explore empirical \tel--\tel\ relations. Although many works have addressed the topic to a greater or lesser extent, none of them had as their main objective the obtaining and discussion of one or more \tel--\tel\ relations; they are always treated as a tool in the analysis of the chemical composition or metallicity of the particular sample of objects observed or compiled in each work. Unlike the previous cases, this paper focuses exclusively on the \tel--\tel\ relations. Our goal is to obtain a set of \tel\ values calculated homogeneously and encompassing all available indicators in the optical spectra of a broad sample of star-forming regions in the Local Universe. To achieve this, we use a careful compilation of high-quality spectra and employ a refined methodology and updated atomic datasets to perform the necessary calculations.

In this paper, we recalculate \tel\ values obtained from diagnostics based on six line ratios of CELs of different ions for a large sample of star-forming regions of the Local Universe, focused primarily on deep high-quality emission-line spectra of Galactic and extragalactic \hii\ regions. With all those data we analyse the behaviour of the \tel--\tel\ relations that can be defined from pairs of those \tel\ indicators. In  
Sect.~\ref{sec:description_sample} we describe the sample of spectra taken from the literature that we have used in this work. All of them have good determinations of, at least, two \tel\ diagnostics. In Sect.~\ref{sec:physical_conditions}, we explain how we compute the physical conditions of each spectrum: its electron density (\nel) and the different \tel\ values that can be obtained from the CELs measured in it. In Sect.~\ref{sec:Te-Te_relations}, we analyse the behaviour of each \tel--\tel\ relation obtained for the different \tel\ pairs selected, showing the results of the linear fits to the relations and the dispersion of the data around the fits. We  discuss the results comparing with the predictions of photoionisation models and with previous results obtained by other authors from observational data. In Sect.~\ref{sec:application} we give some suggestions on the application of the \tel--\tel\ relations obtained. Finally, in Sect.~\ref{sec:conclusions} we summarise our main conclusions.

\section{Description of the sample}
\label{sec:description_sample}

In this study we make use of the DEep Spectra of Ionised REgions Database \citep[DESIRED;][]{MendezDelgado:23b} Extended project \citep[DESIRED-E;][]{MendezDelgado:24b}. DESIRED-E is a compilation of high-quality published spectra of ionised nebulae -- mainly extragalactic \hii\ regions, SFGs, and Galactic \hii\ regions, planetary nebulae and ring nebulae around evolved massive stars -- that feature at least one \tel\ determination based on the following auroral to nebular intensity ratios of CELs: \foiii~$\lambda4363/\lambda 5007$, \fnii~$\lambda5755/\lambda6584$, and/or \fsiii~$\lambda 6312/\lambda9069$. For each spectrum included in DESIRED-E, we collect the extinction-corrected line intensity ratios and their associated uncertainties directly from the source papers and perform a homogeneous analysis. We only consider emission lines with observational errors less than 40\%. Since DESIRED-E is frequently expanded with new spectra, the sample used in this work  corresponds to its state as of January 9, 2025, and contains 3154 spectra. 

\begin{figure}[ht!]
\centering    
\includegraphics[width=\hsize ]{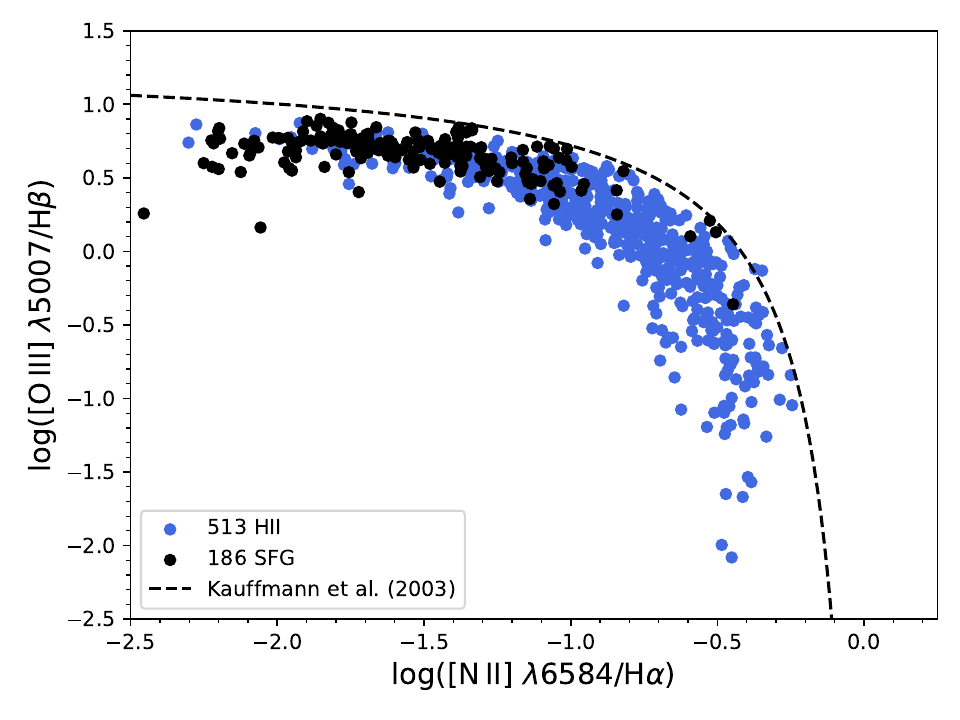}
\includegraphics[width=\hsize ]{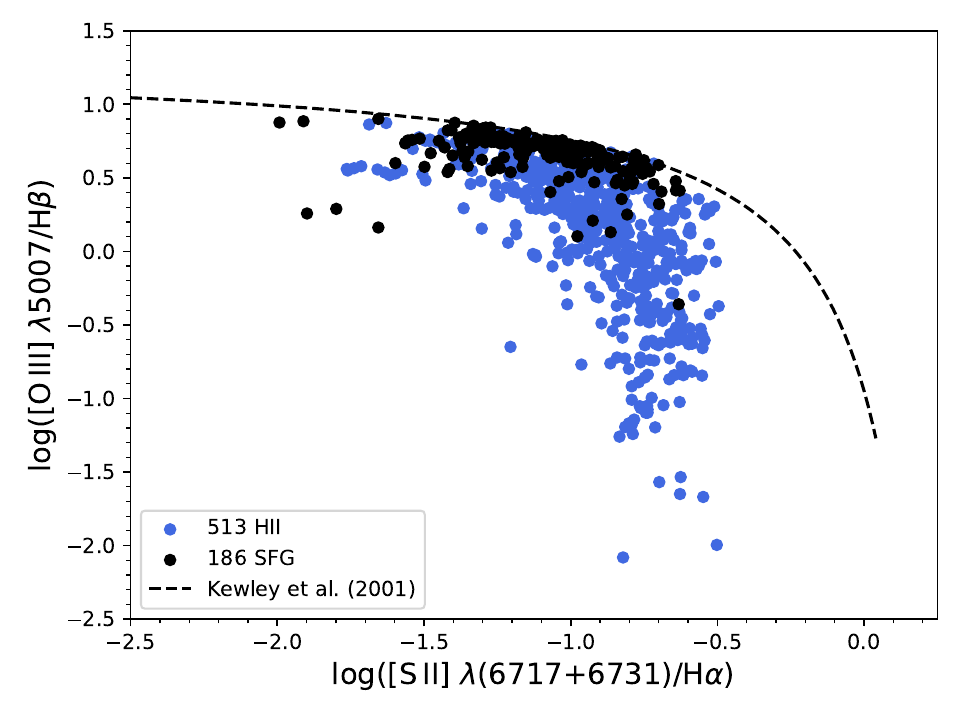}
\caption{Diagrams showing log(\foiii$\lambda5007/$H$\beta$) versus log(\fnii$\lambda6584/$H$\alpha$) (BPT, top) and log(\foiii$\lambda5007/$H$\beta$) versus log(\fsii$\lambda\lambda 6716+31/$H$\alpha$) (bottom) of the sample of spectra of Galactic and extragalactic \hii\ regions (blue dots) and star-forming galaxies (SFGs) (black squares) compiled in DESIRED-E used in this study. The dashed lines in both diagrams represent the empirical relations that have been used to distinguish between star-forming regions and active galactic nuclei (AGNs): that by \citet{Kauffmann:03} in the top diagram and \citet{Kewley:01} in the bottom one.} 
\label{fig:BPT_diagrams}
\end{figure}

 The aim of the present work is to explore the \tel--\tel\ relations in star-forming regions of the local Universe, so we limit the sample to spectra of Galactic and extragalactic \hii\ regions and SFGs of DESIRED-E. To do this, we first filtered the sample by selecting only those spectra whose combinations of the line ratios \foiii~$\lambda5007/$H$\beta$, \fnii~$\lambda6584/$H$\alpha$, and \fsii~$\lambda\lambda 6716+31/$H$\alpha$, are consistent with pure photoionisation, as expected for star-forming regions. In the top panel of Fig.~\ref{fig:BPT_diagrams} we show the diagnostic diagram representing \foiii~$\lambda5007/$H$\beta$ versus \fnii~$\lambda6584/$H$\alpha$, commonly known as Baldwin-Phillips-Terlevich (BPT) diagram \citep{Baldwin:81}. The selected spectra of \hii\ regions (blue dots) and SFGs (black dots) are those located below of the \citet{Kauffmann:03} line, that is  used to distinguish between star-forming regions and active galactic nuclei (AGNs). In the bottom panel of Fig.~\ref{fig:BPT_diagrams} we show the \foiii~$\lambda5007/$H$\beta$ versus \fsii~$\lambda\lambda 6716+31/$H$\alpha$ diagnostic diagram, originally proposed by \citet{Veilleux:87}. In this second diagram, we use the line proposed by \citet{Kewley:01} to separate star-forming regions from AGNs. We use both diagrams and not only the \foiii~$\lambda5007/$H$\beta$ versus   \fnii~$\lambda6584/$H$\alpha$ one -- which is the most commonly used -- to ensure that at least lines of two low-ionisation species are well measured. This is especially important for metal-poor SFGs, whose high-ionisation degree can make it difficult to detect such lines in objects with low surface brightness. 
 
The second filter applied when selecting the spectra is that, from the lines reported in the reference of each spectrum, \tel\ can be calculated with at least two diagnostics of the six available in the optical range: {\tel}(\fnii~$\lambda 5755/\lambda 6584$), {\tel}(\foii~$\lambda\lambda7319+20+30+31/\lambda\lambda 3726+29$), {\tel}(\foiii~$\lambda 4363/\lambda 5007$), {\tel}(\fsii~$\lambda\lambda4069+76/\lambda\lambda 6716+31$), {\tel}(\fsiii~$\lambda 6312/\lambda 9069$), and {\tel}(\fariii~$\lambda5192/\lambda7135$). This  filter was applied to exclude excessively high \tel\ values due to incorrect identification of a given auroral line. To ensure that, we excluded \tel\ values above 20,000 K. This is a reasonable maximum value, taking into account theoretical prescriptions and the values typically observed in local metal-poor SFGs \citep[e.g.][]{Aller:99,PerezMontero:03,Izotov:06,Hagele:08}.

 The total number of DESIRED-E spectra that fulfill all our requirements is 699, of them 513 (73.4\%) correspond to {\hii} regions and 186 (26.6\%) to SFGs. All these spectra have direct determinations of the total O abundance, the proxy of metallicity in ionised nebulae studies. The range of 12+log(O/H) covered by our objects goes from 7.12 to 8.92. In Table A.1 we list the 699 spectra included in the present study, indicating their object name, whether they correspond to {\hii} regions or SFGs and the reference of their published spectra.

\section{Electron density and temperature determinations}
\label{sec:physical_conditions}

We use \textit{PyNeb 1.1.18} \citep{Luridiana:15, Morisset:20} with the \hi~ effective recombination coefficients from \citet{Storey:95} and the atomic dataset for the N, O, S, Cl, Ar and Fe ions presented in Table~\ref{table:atomic_data} to determine {\nel} and {\tel} for each individual spectrum. 

\begin{table*}
    \centering    
    \caption{References for atomic data used for collisionally excited lines.}
    \label{table:atomic_data}
    \begin{tabular}{lcc}
        \hline
        \noalign{\smallskip}
        Ion & Transition probabilities & Collision strengths\\
        \noalign{\smallskip}
        \hline
        \noalign{\smallskip}
            N$^{+}$   &  \citet{Fischer:04} & \citet{Tayal:04}\\
            O$^{+}$   &  \citet{FroeseFischer:04} & \citet{Kisielius:09}\\
            O$^{2+}$  &  \citet{Wiese:96}, \citet{Storey:00} & \citet{Storey:14}\\
            S$^{+}$   &  \citet{Irimia:05} & \citet{Tayal:10}\\
            S$^{2+}$  &  \citet{FroeseFischer:06} & \citet{Grieve:14}\\
            Cl$^{2+}$ &  \citet{Fritzsche:99} & \citet{Butler:89}\\
            Ar$^{2+}$ &   \citet{Mendoza:83b}, \citet{Kaufman:86}  & \citet{Galavis:95}\\
            Ar$^{3+}$ &   \citet{Mendoza:82b}  & \citet{Ramsbottom:97}\\
            Fe$^{2+}$ & \citet{Deb:09}, \citet{Mendoza:23} & \citet{Zhang:96}, \citet{Mendoza:23}\\
        \noalign{\smallskip}
        \hline
    \end{tabular}
\end{table*}

\subsection{Electron densities}
\label{subsec:densities}

The procedure for calculating \nel\ and \tel\ of the sample objects begins using the \textit{getCrossTemDen} routine of \textit{PyNeb} with line intensity ratios sensitive to those quantities. The routine cross-correlates the \nel-sensitive diagnostics \foii~$\lambda 3726/\lambda3729$, \fsii~$\lambda 6731/\lambda6716$, \fcliii~$\lambda 5538/\lambda5518$, \fariv~$\lambda 4740/\lambda4711$, and \ffeiii~$\lambda 4658/\lambda4702$ with the \tel-sensitive ones \fnii~$\lambda 5755/\lambda 6584$, \foii~$\lambda\lambda7319+20+30+31/\lambda\lambda3726+29$, \foiii~$\lambda 4363/\lambda 5007$, \fsii~$\lambda\lambda4069+76/\lambda\lambda6716+31$,  \fsiii~$\lambda 6312/\lambda 9069$, and \fariii~$\lambda 5192/\lambda 7135$ using a Monte Carlo experiment of 100 iterations to propagate uncertainties in line intensities. The number of iterations was chosen as a good compromise between calculation time and convergence of the result. After each convergence, we obtain a set of \nel\ and \tel\ values for each diagnostic along with their associated uncertainties for each individual point of the experiment. The \nel\ value adopted for each diagnostic corresponds to the mean of the 100 individual points weighted by the inverse square of the error. 

After obtaining the mean value of each \nel\ diagnostic, we apply the criteria proposed by \citet{MendezDelgado:23b} to settle a mean \nel\ representative of each spectrum: 
\begin{itemize}
\item If \nel(\fsii) < 100 cm$^{-3}$, we adopt \nel = 100$\pm$100 cm$^{-3}$. 
\item If 100 cm$^{-3} \leq$ \nel(\fsii) $<$ 1000 cm$^{-3}$, we adopt the average of  \nel(\fsii) and \nel(\foii). 
\item If \nel(\fsii) $\geq$ 1000 cm$^{-3}$, we adopt the average of \nel(\fsii), \nel(\foii), \nel(\fcliii), \nel(\ffeiii), and \nel(\fariv). 
\item In cases where a value of \nel\ is not reported or could not be calculated from the data of the source references, we adopt \nel =100$\pm$100 cm$^{-3}$. 
\end{itemize}
The final adopted values of \nel\ for each spectrum used are given in Table A.2.

\subsection{Electron temperatures}
\label{subsec:temperature}

We calculate six \tel\ diagnostics: \tel(\fnii), \tel(\foii), \tel(\foiii), \tel(\fsii), \tel(\fsiii), and \tel(\fariii), using the \textit{getTemDen} routine of \textit{PyNeb} and the adopted representative \nel\ of each spectrum. We also use a Monte Carlo experiment of 100 iterations to propagate uncertainties in \nel\ and line intensities ratios involved in the \tel\ diagnostics. To ensure a good determination of \tel\ for each diagnostic and object, we verify that the flux of the pairs of nebular lines coming from the same upper atomic level used (e.g., \foiii~$\lambda \lambda 5007, 4959$, \fsiii~$\lambda \lambda 9531, 9069$, \fnii~$\lambda \lambda 6584, 6548$) fit with their theoretical predictions \citep[3.00, 2,47, 3.05, respectively; e.g.][]{Storey:00}. We discard any diagnostic when the observed nebular line intensity ratios differ by more than 20\% from the theoretical values. The \fsiii~$\lambda \lambda 9531, 9069$ line ratio tends to differ from the theoretical one due to contamination of telluric absorption bands \citep[see][for a discussion about this issue]{MendezDelgado:24a}. There are a few objects where only one of the lines of the aforementioned pairs of nebular lines is reported. In these cases, we consider the single nebular line observed in the calculation, assuming that the \tel(\fsiii) value thus obtained is valid. The \tel\ values of the different diagnostics obtained for each spectrum are given in Table A.2. Histograms of values of the different \tel\ indicators used in this paper separated by group of objects (\hii\ regions and SFGs) are shown in Fig.~\ref{fig:Median_Tes}. The median values of each \tel\ indicator -- \tel(\fnii), \tel(\foii), \tel(\foiii), \tel(\fsii), \tel(\fsiii), and \tel(\fariii) -- separated for \hii\ regions and SFGs are given in Table~\ref{table:Median_Tes}.

\section{The ${T_{\mathrm{e}}}$-${T_{\mathrm{e}}}$ relations}
\label{sec:Te-Te_relations}

In this section, we present and discuss the \tel--\tel\ relations obtained from our selected sample of DESIRED-E spectra of \hii\ regions and SFGs that satisfy the selection criteria described in Sect.~\ref{sec:description_sample}. Depending on the ionisation potential of the ions involved, we can distinguish \tel\ diagnostics of different ionisation zones within the nebula. \tel(\fnii), \tel(\foii), and \tel(\fsii) are commonly considered \tel\ indicators of the low-ionisation zone, which corresponds, in principle, to the outermost parts in radiation-bounded nebulae. On the other hand, \tel(\fsiii) is usually assumed as representative of the intermediate-ionisation zone, while \tel(\foiii) is considered indicative of the innermost high-ionisation zone of a typical star-forming region \citep[e.g.][]{Berg:20}. A particular case is that of \tel\fariii\ -- of which we have only a few dozen determinations in DESIRED-E -- which, due to the ionisation potential range where we can find Ar$^{++}$ ions, would correspond to a zone overlapping with the S$^{++}$ and O$^{++}$ ones \citep{MendezDelgado:23b}.

With six temperature diagnostics available, many \tel--\tel\ relations can be explored. However, we limit our analysis to those that have been most extensively studied in the literature or are of particular interest, either due to the ions involved or due to possible effects or circumstances that may affect one of the \tel\ diagnostics involved. In this section, we divide the analysis into three subsections. In the first, we examine the relations between the three \tel\ diagnostics for the low-ionisation zone. In the second and third subsections, we present and discuss \tel--\tel\ relations involving \tel(\fsiii) and \tel(\foiii) on the x-axis, respectively.

For each particular \tel--\tel\ relation, we calculate a linear fit to analyse the correlation between both \tel\ diagnostics involved and compare it with the results of previous studies. Following the suggestion of \citet{Hogg:2010}, we use orthogonal distance regression (ODR) technique for the linear fits. This method is especially indicated when both variables are subject to error and their uncertainties are comparable. We note that the $T_{\rm e}-T_{\rm e}$ relations available in the literature have been derived using a variety of fitting techniques. For instance, while we employ an ODR method, other works have used empirical model fits \citep[e.g.,][]{Garnett:92}, standard least-squares polynomials \citep[e.g.,][]{MendezDelgado:23b}, or a hierarchical Bayesian approach like \texttt{linmix} \citep[e.g.,][]{Zurita:21, RickardsVaught:24}. The choice of regression can influence the final slope values. However, as we noted before, we consider the ODR method the most appropriate for this analysis, as it treats uncertainties in both variables symmetrically. We caution that some systematic differences between the slopes reported here and those in previous studies may be driven, in part, by differences in statistical treatment rather than by purely physical variations. An important advantage of the applied ODR is that it provides an invertible linear fit, i.e. the slope of the fit to the \tel(Y)--\tel(X) relation is just the reciprocal value of the \tel(X)--\tel(Y) one.

To obtain the values of the slope and intercept, we perform a Monte Carlo simulation with 10,000 iterations of the fit, using the implementation provided in the \texttt{scipy.ODR} module of SciPy. The final values and uncertainties are estimated as the median and standard deviation of the resulting distributions, respectively. To evaluate the strength and the statistical significance of the correlation between the two diagnostics used in each \tel--\tel\ relation we calculate the Pearson correlation coefficient, $r$, and the $p$-value\footnote{A value of $p$ $>$ 0.05 indicates that the null hypothesis is plausible, i.e. there is not significant linear fit between the variables. On the other hand, when $p$ $\leq$ 0.05 it can be said that there is a significant correlation between both variables.} of each fit. In addition, we calculate the total and intrinsic dispersion of the observational points around the linear fit, $\sigma_{tot}$ and $\sigma_{int}$, respectively \citep{Rogers:21}. We calculate $\sigma_{tot}$ as the standard deviation of the vertical residuals, corrected for degrees of freedom. We then estimate the projected uncertainty in the dependent variable and, with this, construct a reduced $\chi^2$ that includes an additional dispersion $\sigma_{int}$. This $\sigma_{int}$ is defined as the minimum amount that makes the reduced $\chi^2=1$. Thus, $\sigma_{tot}$ measures the actual dispersion of the residuals around the fitted relation, while $\sigma_{int}$ reflects the extra physical dispersion not attributable to measurement errors, both in the y-axis. The parameters of the ODR linear fits of the \tel--\tel\ relations -- as well as the total number of data points, the \tel\ range they cover on the x-axis, $\sigma_{tot}$ and $\sigma_{int}$ -- are presented and discussed in the following subsections and included in Table~\ref{table:ODR_fits}. By providing invertible linear fits, we also calculate the corresponding $\sigma_{tot}$ and $\sigma_{int}$ of the \tel\ indicator that lies on the x-axis of a given \tel--\tel\ relation, calculated from the inverse ODR fit; these values are given in brackets in the last two columns of Table~\ref{table:ODR_fits}.

It is clear that an intrinsic source of scatter in the \tel--\tel\ relations  obtained in this work is whether the nebulae are radiation-bounded or not ---a characteristic that can hardly be verified for the objects in the sample. In principle, radiation-bounded nebulae should develop an ionisation stratification that would also be reflected in an internal temperature gradient. The truncation of this structure in a density-bounded nebula will alter the proportion between the different \tel\ diagnostics, primarily affecting relations involving \tel\ diagnostics of the low- and high-ionisation zones. Furthermore, the truncation of the nebular structure implies a lower fraction of low-ionisation ions, which are typically more efficient at nebular cooling under the standard physical conditions of photoionised nebulae; this will also affect their nominal \tel\ values.

In \cite{MendezDelgado:23b}, we presented initial results on \tel--\tel\ relations based on the initial version of DESIRED, limited to 190 nebular spectra. The present work represents a significant increase in the number of spectra and a considerable advance over \cite{MendezDelgado:23b}. While that study laid important groundwork, our extended sample (DESIRED-E) goes beyond a simple numerical increase; it introduces a population of high-temperature star-forming galaxies that was previously underrepresented. This inclusion has resulted in \tel--\tel\ relations that, in some cases, differ significantly from those of the 2023 study, especially in the high-excitation regime. Therefore, the broader coverage of physical conditions in DESIRED-E provides a more universal calibration than that offered by \cite{MendezDelgado:23b}.

Before describing our results on the \tel--\tel\ relations, we must remember that our study is based on a large compilation of heterogeneous data from the literature. From the papers used, we obtained the line ratios as well as their associated errors. As is well known, the criteria for estimating errors can differ significantly from one study to another, so the same line ratio could have a different nominal error in different studies. This makes it difficult, for example, to conduct a comparative study of the effect that different error values in the intensity of auroral lines might have on the \tel--\tel\ relations. Furthermore, the treatment of errors in flux calibration and reddening corrections is often different in different works and may introduce correlations between different temperature measurements of the spectral lines. However, since, as mentioned, our database is a meta-analysis of heterogeneous bibliographic sources, the raw calibration data and the covariance matrices necessary to quantify these correlations are not available. While these systematic effects may contribute to the observed dispersion in our relationships, the use of a large and diverse sample of objects might help to average out the calibration biases specific to each study.

\subsection{Relations involving diagnostics of the low-ionisation zone}
\label{subsec:low_ion}

The three \tel\ diagnostics of the low-ionisation zone that can be derived from the optical spectra compiled in DESIRED-E are \tel(\fnii), \tel(\foii), and \tel(\fsii). In our sample, \tel(\fnii) can be determined for 371 spectra (350 \hii\ regions and 21 SFGs). For \tel(\foii), there are 558 spectra (435 \hii\ regions and 123 SFGs),  and for \tel(\fsii), 487 spectra (361 \hii\ regions and 126 SFGs). The comparison of the numbers corresponding to \hii\ regions and SFGs, clearly indicates the difficulty of detecting and measuring auroral lines from low-ionisation species in SFGs, which are generally of low metallicity and therefore show a high degree of ionisation. The difficulty is especially pronounced in the case of \fnii~$\lambda$5755 because metal-poor SFGs almost invariantly show a lower N/O ratio, making that line extremely weak in such objects. 

Photoionisation models predict that the three \tel\ diagnostics representative of the low-ionisation zone follow the relation \tel(\fnii) $\approx$ \tel(\foii) $\approx$ \tel(\fsii) \citep{Campbell:86, Garnett:92, Pagel:92, Thuan:95, Izotov:07, MendezDelgado:23b}. However, this behaviour is not always found consistently recovered in the \tel--\tel\ distributions based on observational data, as we will see below. 

\begin{figure}[ht!]
\centering    
\includegraphics[width=\hsize ]{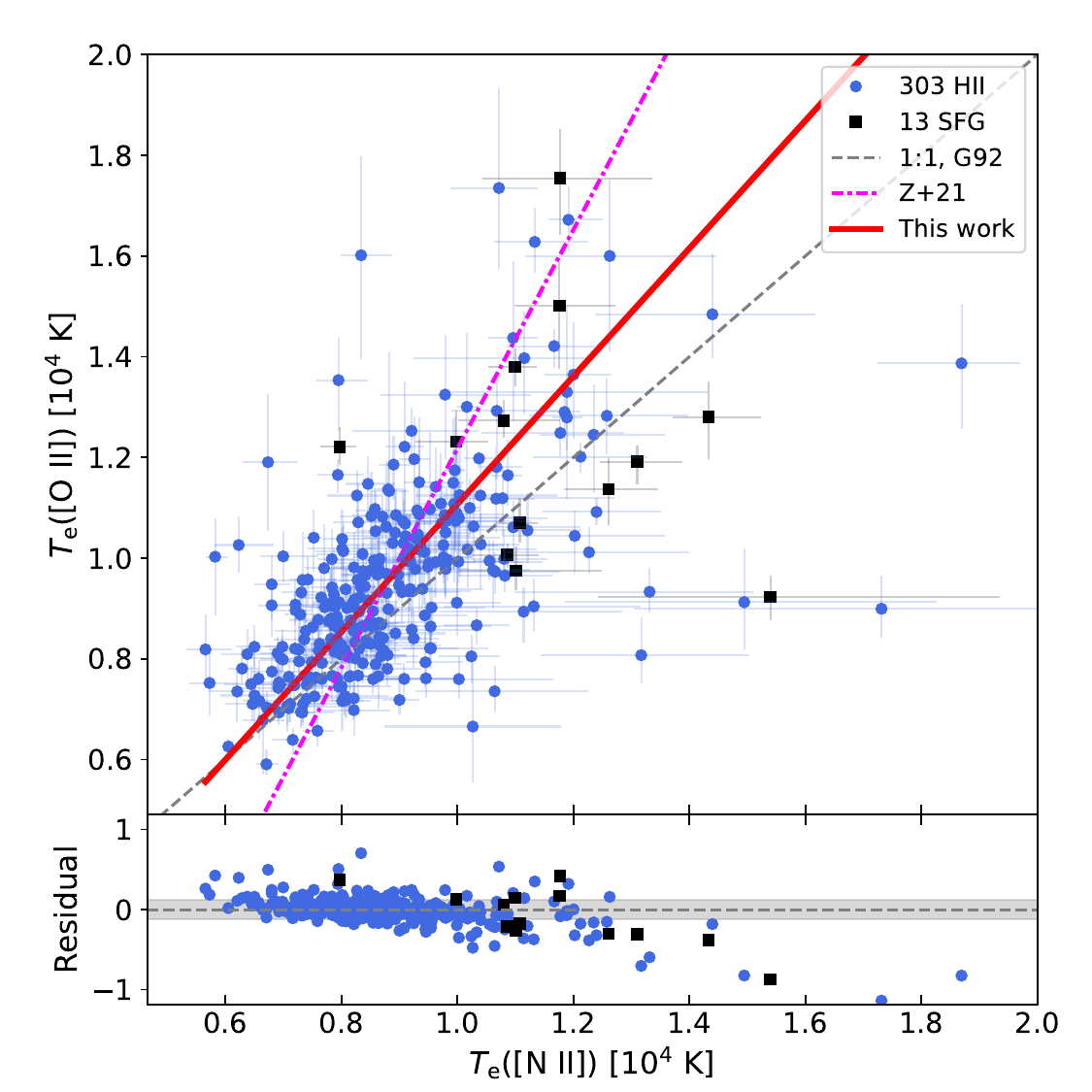}
\includegraphics[width=\hsize ]{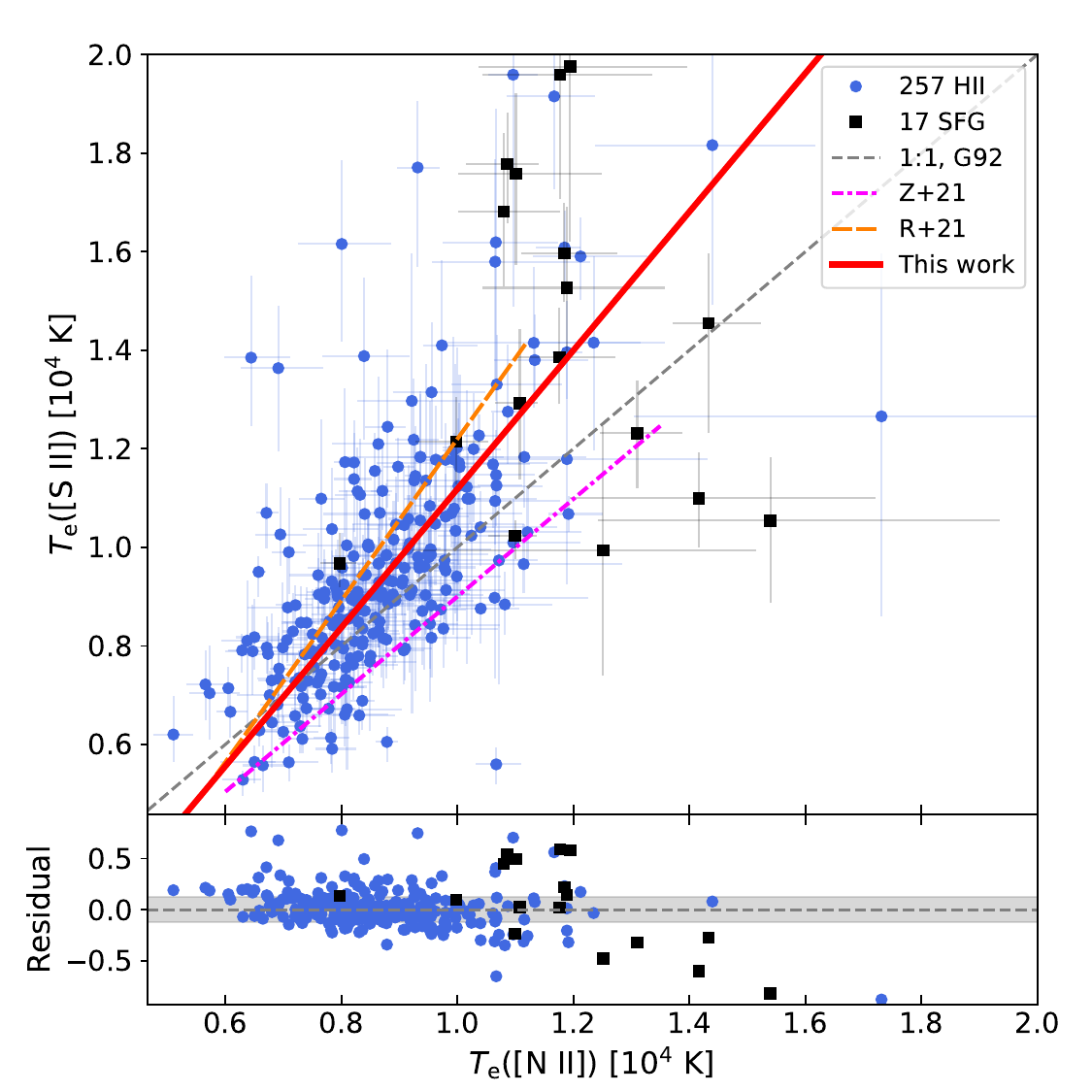}
\caption{\tel(\foii)--\tel(\fnii) (top) and \tel(\fsii)--\tel(\fnii) (bottom) relations obtained for our DESIRED-E sample. In both panels, blue dots correspond to \hii\ regions and black squares to SFGs, while the red continuous lines represent the ODR linear fits to the data. The grey dashed line shows the 1:1 relation that coincides with the  approximate predictions of photoionisation models by \citet{Garnett:92}, and the magenta dashed line represents the linear fit obtained by \citet{Zurita:21}. In the bottom panel, the orange dashed line represents the fit obtained by \citet{Rogers:21}
. The fits obtained from the literature are only shown covering the \tel\ range from  the corresponding observations on each reference. The rectangular inset below each plot shows the vertical residuals to the ODR linear fit in units of $10^4$ K, with the grey band showing the total dispersion ($\sigma_{tot}$) of the data around the fit.} 
\label{fig:TO2-TN2_TS2-TN2}
\end{figure}

\begin{figure}[ht!]
\centering    
\includegraphics[width=\hsize ]{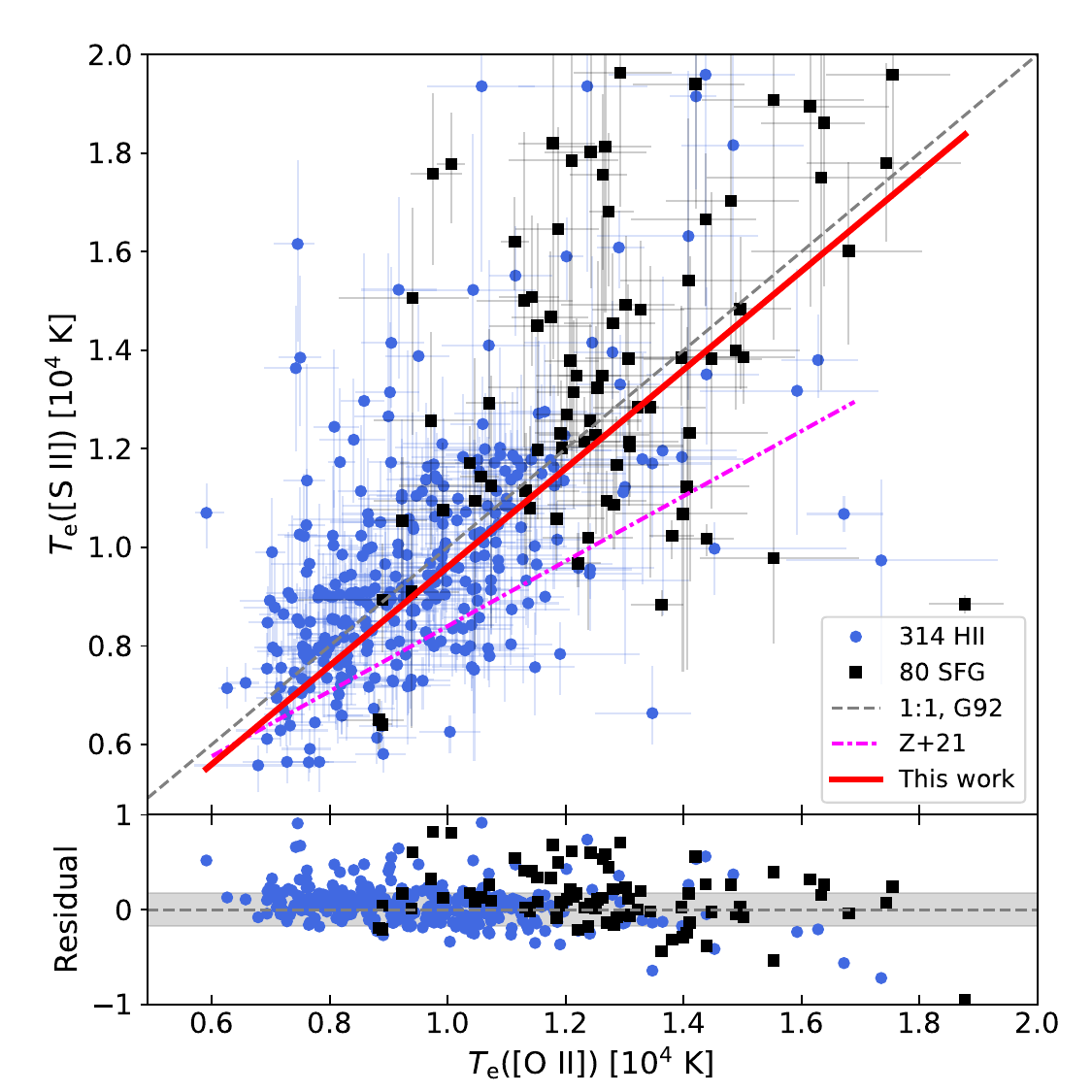}
\caption{Same as the upper panel of Fig.~\ref{fig:TO2-TN2_TS2-TN2} but for the \tel(\fsii)--\tel(\foii) relation.} 
\label{fig:TS2-TO2}
\end{figure}

In the top panel of Fig.~\ref{fig:TO2-TN2_TS2-TN2} we show the \tel(\foii)--\tel(\fnii) relation obtained for 316 DESIRED-E spectra (303 \hii\ regions and 13 SFGs) and the ODR linear fit to the data points is: 
\begin{equation}
\label{eq:TO2-TN2}
    T_{\rm e}(\text{\foii}) = (1.269\pm0.039)\times T_{\rm e}(\text{\fnii}) - (1620\pm320)~{\rm K}.
\end{equation}
The fit parameters given in Table~\ref{table:ODR_fits} indicate that the correlation is clear but moderate ($r$ = 0.59) and statistically significant attending to its extremely small $p$-value.\footnote{The p-values we found in all the temperature relationships studied in this work are extremely low -- between 4.9$\times10^{-49}$ to 6.2$\times10^{-15}$, indicating that all of them are statistically significant. We do not refer to the value of this parameter again in the rest of the \tel--\tel\ relations, to avoid repetition.} As it is  shown in Table~\ref{table:ODR_fits}, the $\sigma_{tot}$(\tel(\foii)) of the observational data about the linear fit is 1940 K and the $\sigma_{int}$(\tel(\foii)) is 1160 K. The total and intrinsic dispersions in \tel(\fnii) are $\sigma_{tot}$(\tel(\fnii)) = 1530 K and $\sigma_{int}$(\tel(\fnii)) = 920 K, respectively.  The slope of the linear fit to \tel(\foii)--\tel(\fnii) is greater than one and therefore away from the predictions of the photoionisation models. Other fits of observational data of this relation or its inverse from the literature give larger values of the slope -- of about 2.0 -- as those by \citet[][132 \hii\ regions]{Rogers:21}, \citet[][196 \hii\ regions]{Zurita:21} or \citet[][132 \hii\ regions]{RickardsVaught:24}, but also smaller -- about 1.0 -- as in \citet[][115 \hii\ regions]{Berg:20} or \citet[][290 SFGs]{Scholte:26}. In the inset at the bottom of each panel in Fig~\ref{fig:TO2-TN2_TS2-TN2} (as well as in figs.~\ref{eq:TS2-TO2} through \ref{fig:TS3-TO3_TAr3-TO3}) we include the residuals to the linear ODR fit along with a grey band indicating its  corresponding $\sigma_{tot}$ value. In the inset it is clear that \tel(\fnii) > \tel(\foii) when \tel(\fnii) $>$ 12,000 K. The residuals shown in the insets of Figures~\ref{fig:TO2-TN2_TS2-TN2}--\ref{fig:TS3-TO3_TAr3-TO3} are defined as the vertical difference between the observed and predicted values, $\Delta T_{\rm e} = T_{\rm e}(\text{obs}) - T_{\rm e}(\text{fit})$. It is important to note that while the fits are calculated using an ODR technique (which minimizes the perpendicular distance to the regression line), the residuals are presented as linear offsets in the ordinate in units of $10^4$ K. This approach allows for a direct assessment of the uncertainty introduced when one temperature diagnostic is used to estimate another.

\begin{table*}
    \centering
    \caption{Parameters of the ODR linear fits to  \tel--\tel\ relations. Values of $\sigma$ in parenthesis correspond to the inverse relation (see third paragraph of Sect.~\ref{sec:Te-Te_relations})}
    \label{table:ODR_fits}
    \begin{tabular}{cccccccc}
        \hline
        \noalign{\smallskip}
        & & \tel\ range & & Intercept & & $\sigma_{tot}$ & $\sigma_{int}$ \\
        \tel--\tel & No. & (K) & Slope & (K) & $r$ & (K) & (K) \\
        \noalign{\smallskip}
        \hline
        \noalign{\smallskip}
        \multicolumn{8}{c}{Relations involving diagnostics of the low-ionisation zone}\\
        \tel(\foii)--\tel(\fnii) & 316 & 5660--18690 & $1.269\pm0.039$ & $-1620\pm320$ & 0.59 &  1940 (1530) & 1160 (920) \\
        \tel(\fsii)--\tel(\fnii) & 274 & 5100--17310 & $1.406\pm0.060$ & $-2880\pm480$ & 0.65 &  2140 (1520) & 1280 (910) \\
        \tel(\fsii)--\tel(\foii) & 394 & 5910--18770 & $1.001\pm0.044$ & $-400\pm380$ & 0.64 &  2430 (2430) & 1740 (1740) \\
        \multicolumn{8}{c}{Relations involving \tel(\fsiii)}\\
        \tel(\fsii)--\tel(\fsiii) & 257 & 5420--19850 & $0.705\pm0.025$ & $2760\pm210$ & 0.63 &  2090 (2970) & 1410 (2010) \\       
        \tel(\fnii)--\tel(\fsiii) & 214 & 4930--16000 & $0.807\pm0.022$ & $1890\pm180$ & 0.75 &  1240 (1540) & 650 (800) \\
        \tel(\foii)--\tel(\fsiii) & 303 & 4930--19850 & $0.812\pm0.020$ & $2390\pm170$ & 0.68 &  1770 (2140) & 1480 (1800) \\    
        \multicolumn{8}{c}{Relations involving \tel(\foiii)}\\
        \tel(\foii)--\tel(\foiii) & 380 & 7320--19860 & $0.523\pm0.017$ & $5120\pm170$ & 0.63 &  1830 (3500) & 1440 (2770) \\ 
        \tel(\fnii)--\tel(\foiii) & 204 & 7290--15730 & $0.763\pm0.032$ & $2530\pm280$ & 0.70 &  1360 (1780) & 790 (1040) \\ 
        \tel(\fsiii)--\tel(\foiii) & 224 & 7320--19540 & $0.953\pm0.015$ & $580\pm150$ & 0.79 &  1510 (1580) & 1280 (1340) \\ 
        \tel(\fariii)--\tel(\foiii) & 32 & 7780--15600 & $0.911\pm0.069$ & $540\pm610$ & 0.93 &  890 (970) & 160 (180) \\
        \noalign{\smallskip}
        \hline
    \end{tabular}
\end{table*}

In the bottom panel of Fig.~\ref{fig:TO2-TN2_TS2-TN2} we show the \tel(\fsii)--\tel(\fnii) relation obtained from 274 DESIRED-E spectra  (257 \hii\ regions and 17 SFGs). The corresponding ODR linear fit to these data points is: 
\begin{equation}
\label{eq:TS2-TN2}
    T_{\rm e}(\text{\fsii}) = (1.406\pm0.060)\times T_{\rm e}(\text{\fnii}) - (2880\pm480)~{\rm K}. 
\end{equation}

The Pearson coefficient of this fit is $r$ = 0.65, indicating that the correlation is moderate-strong. For \tel(\fsii), the total and intrinsic dispersions are 2140 K and 1280 K; and 1520 K and 910 K for \tel(\fnii), respectively. Previous works give different values of the slope of the linear fit to the \tel(\fsii)--\tel(\fnii) relation. From spectra of extragalactic \hii\ regions, \citet{Rogers:21} obtain a value of 1.64 from 123 objects;  \citet{Zurita:21} give a slope of 0.99 from a compilation of 160 \hii\ regions and \citet{RickardsVaught:24}, using 189 objects, obtain a value of 0.85, the lowest slope of all determinations available. Finally, \citet{Scholte:26} obtain 1.02 from a sample of 87 SFGs. 

The resulting \tel(\fsii)--\tel(\foii) relation obtained from our 394 spectra (314 \hii\ regions and 80 SFGs) is presented in Fig.~\ref{fig:TS2-TO2}. This relation includes significantly more spectra of SFGs than the two previous ones that involve \tel(\fnii). This is due to the paucity of measurements of the \fnii$\lambda$5755 line, especially faint at the low metallicities common in SFGs. The ODR linear fit to the data points shown in the figure is: 
\begin{equation}
\label{eq:TS2-TO2}
    T_{\rm e}(\text{\fsii}) = (1.001\pm0.044)\times T_{\rm e}(\text{\foii}) - (400\pm380)~{\rm K}. 
\end{equation}

Considering its Pearson coefficient, this linear fit  indicates a moderate-strong correlation ($r$ = 0.64). As we see in Eq.~\ref{eq:TS2-TO2}, the slope of the linear fit is entirely consistent with unity within the errors. However, the fit line shows a slight offset from the 1:1 line; \tel(\foii) systematically tends to be about 400 K higher than \tel(\fsii). A very similar slope was obtained by \citet{MendezDelgado:23b} for a much limited sample of 39 spectra and  using the same methodology.  However, our fit differs significantly from that obtained by \citet{Zurita:21} from a compilation of data for 175 extragalactic \hii\ regions, who find a slope of 0.66. \citet{Scholte:26} obtain an intermediate value of 0.79, based on a sample of 365 SFGs. The total and intrinsic dispersions of the data about our linear fit are larger than those of the \tel--\tel\ relations discussed so far, 2430 K and 1740 K, respectively. Coincidentally, the same values for \tel(\fsii) and \tel(\foii), being $\sigma_{int}$ especially larger in this case with respect to the other \tel--\tel\ relations discussed in this subsection.   

The comparison of Figs.~\ref{fig:TO2-TN2_TS2-TN2} and \ref{fig:TS2-TO2} and the parameters of the linear fits given in Table~\ref{table:ODR_fits} indicate that, while the \tel(\foii)--\tel(\fnii) and \tel(\fsii)--\tel(\fnii) relations show a slope greater than one and very similar to each other, the \tel(\fsii)--\tel(\foii) one shows a slope consistent with unity. This is exactly the same trend reported by \citet{MendezDelgado:23b} for a much smaller subsample -- between 30 and 39 spectra -- of the objects presented in this work. \tel(\foii) and \tel(\fsii) tend to be higher than \tel(\fnii) in most of the range of variation of \tel(\fnii), with a difference that increases toward higher \tel(\fnii), a trend that has also been reported in previous studies \citep{Pilyugin:09,Bresolin:09,Esteban:09,Rogers:21}. \citet{MendezDelgado:23b} carried out an exhaustive analysis to explain the aforementioned behaviour in the subsection 6.1 of their paper, so we do not repeat it here and only mention the most relevant results of such analysis. Those authors list the main causes that have been proposed to explain the discrepancies between \tel(\fnii), \tel(\foii) and \tel(\fsii), that can be classified as observational effects (incorrect flux calibration or reddening correction, telluric contamination, blends with other faint lines) or other physical reasons (mismatch between the N$^+$, O$^+$ and S$^+$ zones, recombination contribution to the auroral CELs, \tel\ fluctuations or variations in the spatial distribution of \nel\ inside the nebula). \citet{MendezDelgado:23b} conclude that density inhomogeneities must certainly be the main responsible for the trends of \tel(\foii) or \tel(\fsii) with respect to \tel(\fnii). As shown in figure 9 of \citet{MendezDelgado:23b}, the line ratios used to determine \tel(\foii) and \tel(\fsii) have a strong but very similar dependence on \nel\ when it is in the range from 100 to 10,000 cm$^{-3}$ -- the common density range in \hii\ regions and SFGs in the local Universe -- while those of \tel(\fnii) begin to be sensitive to \nel\ at values above that range. In the presence of \nel\ inhomogeneities in a nebula, the common indicators \nel(\fsii) and \nel(\foii) are biased towards the zones of lower values of \nel. Using biased density values -- lower than the true ones -- \tel(\foii) and \tel(\fsii) would tend to be larger than \tel(\fnii). 

The high \nel-dependence of \tel(\foii) and \tel(\fsii) indicators means that, in addition to being able to generate trends with respect to \tel(\fnii), it increases the intrinsic dispersion of the \tel--\tel\ relations in which they are used. This effect has been discussed several times in the literature. For example, \citet{Vermeij:02} or \citet{Bresolin:09b} estimate that an uncertainty of 50–100 cm$^{-3}$ in \nel\ can in turn produce uncertainties in \tel(\foii) of the order of 350–500 K, and somewhat lower in the case of \tel(\fsii). \citet{PerezMontero:03}, in their observed \tel(\fsii)--\tel(\foii) and \tel(\foii)--\tel(\foiii) distributions (their figures 3 and 4) introduce curves showing sequences of models computed with \nel\ values of 10, 100, and 500 cm$^{-3}$, which seem sufficient to account for their large scatter. These sequences were later used and discussed by \citet{Hagele:06,Hagele:08} and \citet{ArellanoCordova:20a}. The aforementioned effect is reflected in the large $\sigma_{int}$ values of the \tel--\tel\ relations involving \tel\ indicators of the low ionisation zone shown in Table~\ref{table:ODR_fits}. Whenever \tel(\foii) or \tel(\fsii) appear, the intrinsic dispersion increases, and by a similar proportion. However, all relationships involving \tel(\fnii) consistently exhibit lower intrinsic dispersions. This result supports  one of the conclusions of the analysis of  \citet{MendezDelgado:23b}, in which it is recommended that the use of \tel(\foii) and \tel(\fsii) should be avoided to calculate abundances of low-ionisation ions when \tel(\fnii) is available, due to the greater reliability of the latter indicator. Unfortunately, as has been mentioned on several occasions, \tel(\fnii) is very difficult to obtain at the low metallicities typical of SFGs. This is clearly reflected in \citet{Scholte:26}; from almost 50,000 SFGs with z $<$ 0.96 from the second release of DESI data\footnote{These data are still not public at the moment of the writing of this paper.}, they only have 340 SFGs (0.5\% of the sample) with  \tel(\fnii) determinations.

\subsection{Relations involving ${T_{\mathrm{e}}}$\rm{([S\,\textsc{iii}])}}
\label{subsec:rel_S3}

\tel(\fsiii) is the only diagnostic available in the optical spectra of star-forming regions that is commonly considered representative of the \tel\ of the intermediate-ionisation zone \citep[e.g.][]{Garnett:92,Berg:20}. The total number of selected DESIRED-E spectra with determinations of this \tel\ diagnostic is 345, of which 282 correspond to \hii\ regions and 63 to SFGs. In this subsection we discuss the results for the \tel--\tel\ relations involving \tel(\fsiii) as the second element of the relation (the x-a in the graphical representations). We consider \tel(\fsii)--\tel(\fsiii), \tel(\fnii)--\tel(\fsiii), and \tel(\foii)--\tel(\fsiii) relations, although for the latter we only provide its parameters in Table~\ref{table:ODR_fits}.

\begin{figure}[ht!]
\centering    
\includegraphics[width=\hsize ]{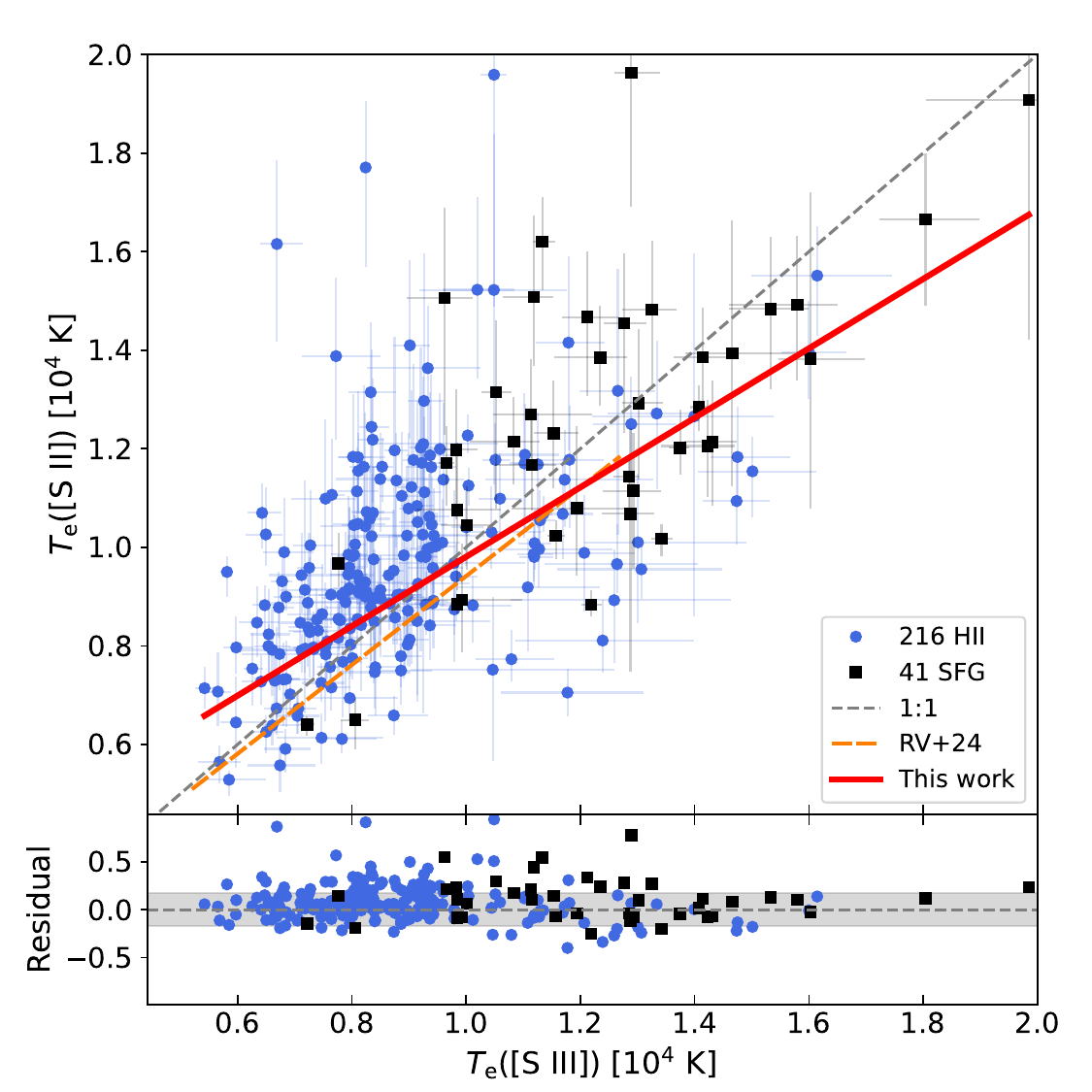}
\includegraphics[width=\hsize ]{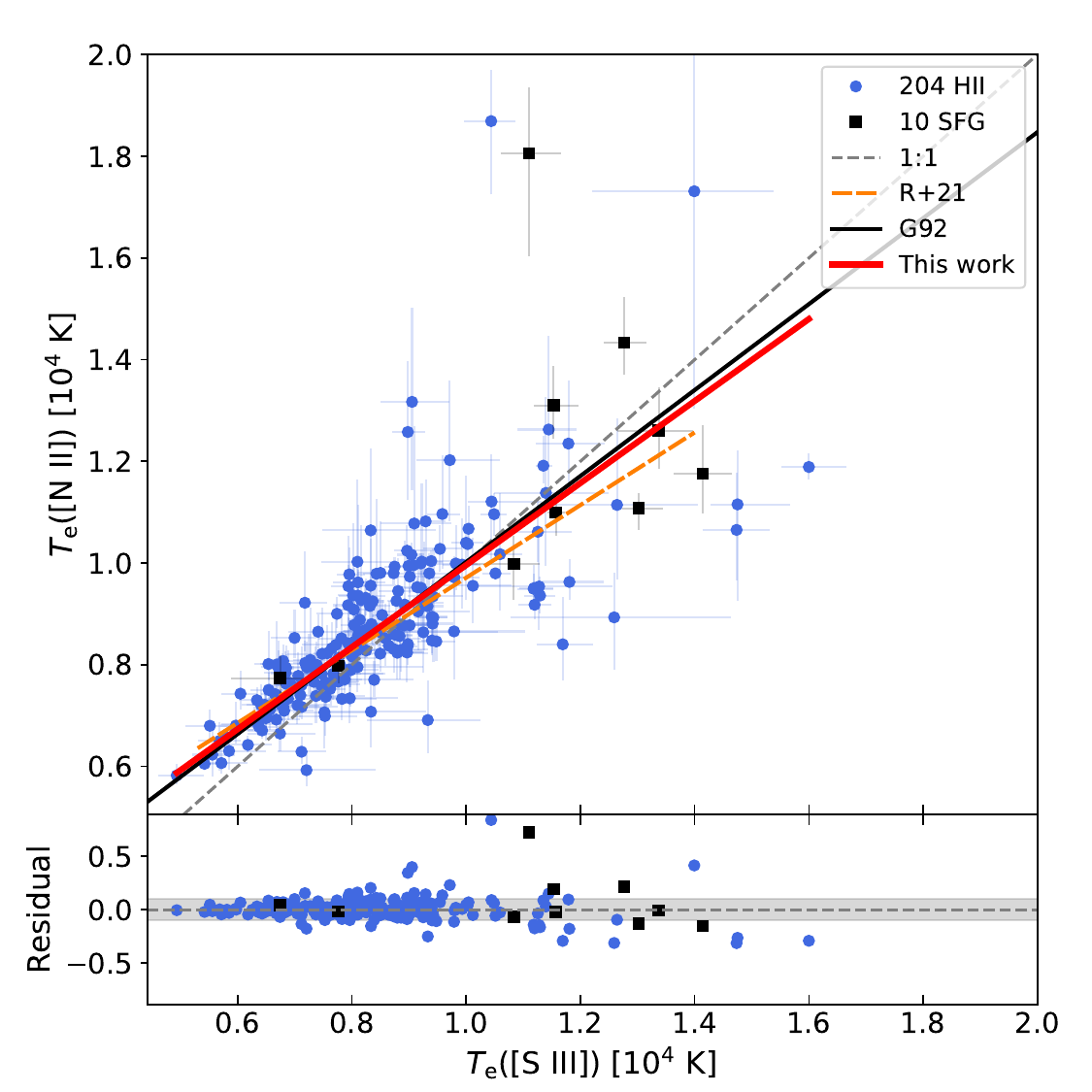}
\caption{\tel(\fsii)--\tel(\fsiii) (top) and \tel(\fnii)--\tel(\fsiii) (bottom) relations obtained for our DESIRED-E sample. The red continuous lines represent the ODR linear fits to the data. The grey dashed lines show the 1:1 relation. The orange dashed lines represent the linear fits to the observational data of \citet{RickardsVaught:24} and \citet{Rogers:21} in the top and bottom panels, respectively. The black continuous line in the bottom panel represents the linear fit obtained from photoionisation models by \citet{Garnett:92}. The rectangular inset below each plot shows the residuals to the ODR linear fit, with the grey band showing the total dispersion ($\sigma_{tot}$) of the data around the fit.} 
\label{fig:TS2-TS3_TN2-TS3}
\end{figure}

The \tel(\fsii)--\tel(\fsiii) relation has received little attention in the literature, although it may be of interest for analyzing how the temperatures of two different nebular zones compare when two ionisation states of the same element are involved. Later, in Section~\ref{subsec:rel_O3}, we discuss the \tel(\foii)--\tel(\foiii) relation, which is an analogous case but has been studied much more extensively. In the top panel of Fig.~\ref{fig:TS2-TS3_TN2-TS3} we represent the \tel(\fsii)--\tel(\fsiii) relation we obtain from 257 DESIRED-E spectra  (216 \hii\ regions and 41 SFGs). The ODR linear fit to the data is: 
\begin{equation}
\label{eq:TS2-TS3}
    T_{\rm e}(\text{\fsii}) = (0.705\pm0.025)\times T_{\rm e}(\text{\fsiii}) + (2760\pm210)~{\rm K}. 
\end{equation}

The correlation between the two \tel\ indicators is moderate-strong and statistically significant, with a Pearson coefficient of 0.63. The $\sigma_{tot}$ and $\sigma_{int}$ values about the linear fit to this relation for \tel(\fsii) are 2090 K and 1410 K, respectively (2970 K and 2010 K for \tel(\fsiii), respectively). There is no published work that includes an explicit value for the slope of the \tel(\fsii)--\tel(\fsiii) relation predicted by photoionisation models. \citet{RickardsVaught:24} present an invertible linear ODR fit to observational values of the \tel(\fsiii)--\tel(\fsii) relation for a sample of 123 extragalactic \hii\ regions (see Fig.~\ref{fig:TS2-TS3_TN2-TS3}) obtaining a slope of 0.90, slightly higher than ours. On the other hand, \citet{Scholte:26} obtain 0.705 -- a slope quite similar to ours -- from measurements of 258 SFGs.

Unlike \tel(\fsii)--\tel(\fsiii), the \tel(\fnii)--\tel(\fsiii) relation has been widely discussed in the literature \citep[e.g.][]{Berg:15,Berg:20,Croxall:16,Rogers:21,Zurita:21,MendezDelgado:23b, RickardsVaught:24}. In the bottom panel of Fig.~\ref{fig:TS2-TS3_TN2-TS3} we represent the \tel(\fnii)--\tel(\fsiii) relation we obtain from 214 DESIRED-E spectra  (204 \hii\ regions and only 10 SFGs). The ODR linear fit to the data is: 
\begin{equation}
\label{eq:TN2-TS3}
    T_{\rm e}(\text{\fnii}) = (0.807\pm0.022)\times T_{\rm e}(\text{\fsiii}) + (1890\pm180)~{\rm K}. 
\end{equation}

Fig.~\ref{fig:TS2-TS3_TN2-TS3} illustrates the remarkably tight correlation of the \tel(\fnii)--\tel(\fsiii) relation -- especially for \tel(\fsiii) < 9000 K -- with a small dispersion. In fact the values of $\sigma_{tot}$ and $\sigma_{int}$ of this relation for \tel(\fnii) are 1240 K and 650 K, respectively -- 1540 K and 800 K for \tel(\fsiii) --, one of the lowest $\sigma_{int}$ values found in this study. The Pearson coefficient of the fit is 0.75, one of the highest values of all the \tel--\tel\ relations studied. Our slope of 0.807 is just between the values of $~$0.7 and 0.92 obtained by \citet{Garnett:92} and \citet{MendezDelgado:23b}, respectively, from photoionisation models, and somewhat larger than the slopes obtained by \citet{Rogers:21}, \citet{Zurita:21} or \citet{RickardsVaught:24} of 0.69, 0.71, and 0.74, respectively, from observational data. All those authors -- as well as \citet{Berg:15,Berg:20}, \citet{Croxall:16} or \citet{MendezDelgado:23b} -- highlight the strong correlation and small dispersion that this relation shows. Only \citet{Scholte:26} obtain a slope significantly larger, 1.01, from data of 184 SFGs. 

\subsection{Relations involving ${T_{\mathrm{e}}}$\rm{([O\,\textsc{iii}])}}
\label{subsec:rel_O3}

Of the six \tel\ diagnostics we have available, \tel(\foiii) is the one considered representative of the high-ionisation zone in typical star-forming regions \citep[e.g.][]{Garnett:92,Berg:20}. We have 498 spectra with determinations of \tel(\foiii) in our sample, 315 of which are \hii\ regions and 183 SFGs. In this subsection we discuss the results for the \tel--\tel\ relations involving \tel(\foiii) as the second element -- x-axis -- of the relation, in our case \tel(\foii)--\tel(\foiii), \tel(\fnii)--\tel(\foiii), \tel(\fsiii)--\tel(\foiii), and \tel(\fariii)--\tel(\foiii). 

\begin{figure}[ht!]
\centering    
\includegraphics[width=\hsize ]{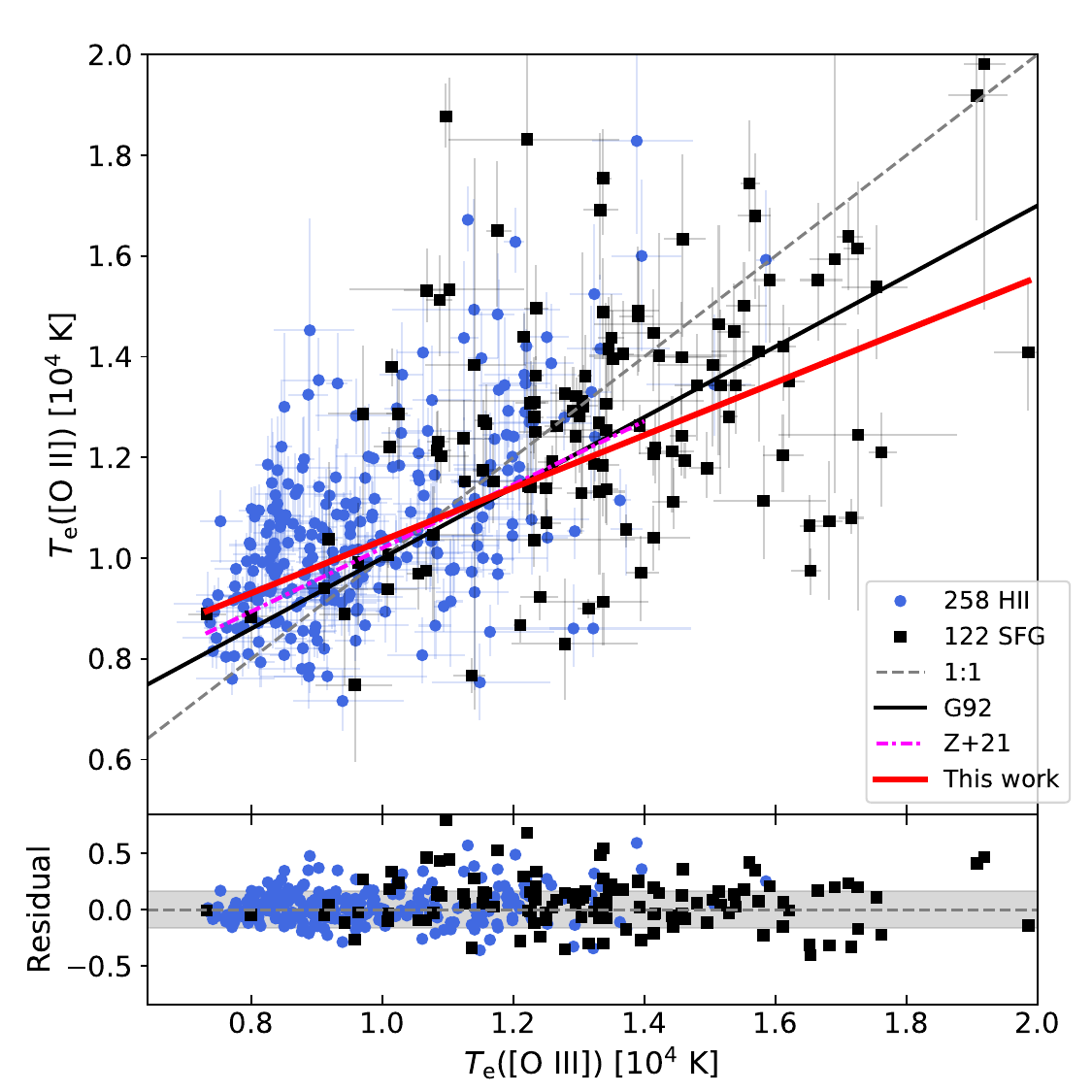}
\includegraphics[width=\hsize ]{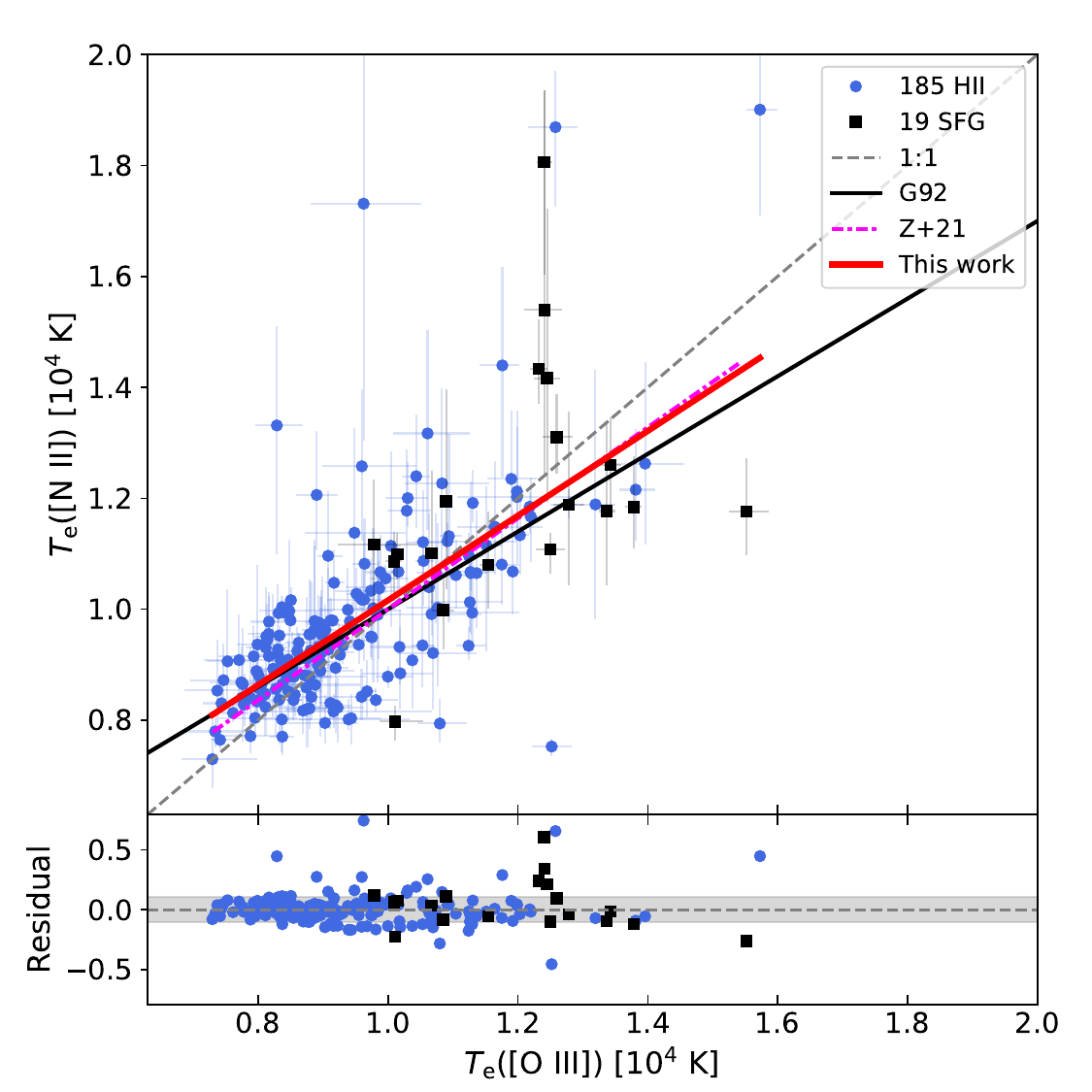}
\caption{\tel(\foii)--\tel(\foiii) (top) and \tel(\fnii)--\tel(\foiii) (bottom) relations obtained for our DESIRED-E sample. The red continuous lines represent the ODR linear fits to the data. The grey dashed lines show the 1:1 relation. The magenta dotted-dashed lines represent the linear fits to the observational data of \citet{Zurita:21}. The black continuous lines represent the linear fits obtained from photoionisation models by \citet{Garnett:92}. The rectangular inset below each plot shows the residuals to the ODR linear fit, with the grey band showing the total dispersion ($\sigma_{tot}$) of the data around the fit.} 
\label{fig:TN2-TO3_TO2-TO3}
\end{figure}

In the top panel of Fig.~\ref{fig:TN2-TO3_TO2-TO3} we show the \tel(\foii)--\tel(\foiii) relation we obtain from 380 DESIRED-E spectra, of which 258 correspond to \hii\ regions and 122 to SFGs. The distribution of the data points shows a large dispersion and its ODR linear fit is: 
\begin{equation}
\label{eq:TO2-TO3}
    T_{\rm e}(\text{\foii}) = (0.523\pm0.017)\times T_{\rm e}(\text{\foiii}) + (5120\pm170)~{\rm K}. 
\end{equation}

This is the fit with the slope value furthest from unity of all the \tel--\tel\ relations studied. In any case, it shows a strong-moderate ($r$ = 0.63) correlation. It is well known in the literature that the \tel(\foii)--\tel(\foiii) relation in \hii\ regions exhibits a very high dispersion. Even early works such as that of \citet{Kennicutt:03} claimed that the two \tel\ diagnostics were uncorrelated. In our case, the $\sigma_{tot}$ and $\sigma_{int}$ values of this relation for \tel(\foii) are quite large, 1830 K and 1440 K, respectively, and even larger for \tel(\foiii) (3500 K and 2770 K, respectively). In Section~\ref{subsec:low_ion}, following the conclusions of the analysis of \citet{MendezDelgado:23b}, we indicated that \nel\ inhomogeneities should be the main responsible for the trends and large dispersions of \tel(\foii) or \tel(\fsii) with respect to other \tel\ indicators less sensitive to collisional de-excitation at the typical \nel\ values of star-forming regions. Although the spatial variations in \nel\ may be the most important, there are other physical phenomena that can also contribute to the larger dispersion of the \tel(\foii) diagnostic depending on the characteristics of the nebulae and the quality of the observations. The auroral \foii~$\lambda\lambda7319+20+30+31$ lines can be affected by recombination (dielectronic plus radiative) contribution from O$^{++}$ ions \citep{Rubin:86,Liu:01}. The increase in intensity of the red \foii\ auroral lines due to recombination exhibits a complex and contradictory behaviour. On the one hand, it increases as the \tel\ decreases, and therefore as the metallicity of the object increases \citep{Stasinska:05}. This is because recombination  coefficients are larger at lower \tel. On the other hand, recombination is more significant the higher the proportion of O$^{++}$ present, and it also depends on the degree of ionisation of the nebula, which, in turn, increases at lower metallicity. Using photoionisation models, \citet{MendezDelgado:23b} find that the recombination contribution -- if present -- is clearly higher in the case of \tel(\foii) than in \tel(\fnii) and that the difference between the two \tel\ diagnostics would be correlated with the intensity of the recombination lines of the V1 multiplet of \oii, something that is not observed, although for a rather limited sample of objects due to the difficulty of measuring the extremely faint \oii\ lines. This leads \citet{MendezDelgado:23b} to dismiss a significant contribution of recombination to \tel(\foii) in their sample. In any case, we cannot rule out that this phenomenon could be present in a non-negligible way in some objects -- especially in highly ionised ones -- and thus  contribute to increase the scatter of the \tel(\foii) distribution at higher \tel\ values.

As also discussed in Section~\ref{subsec:low_ion}, the calculation of \tel(\foii) is very sensitive to several observational problems that are difficult to quantify and sometimes even to identify. Several authors have detailed those observational problems that arise mainly from the large wavelength distance that separates the lines from whose ratio the \tel(\foii) diagnostic is based --  \foii~$\lambda\lambda 3726+29$ and \foii~$\lambda\lambda7319+20+30+31$ \citep[e.g.][]{Kennicutt:03, Rodriguez:20, MendezDelgado:23b}. With such a long wavelength baseline, any error in flux calibration or  reddening correction of the optical spectra amplifies the uncertainty of the diagnostic line ratio and hence the final \tel(\foii) error. Finally, the contamination by telluric spectral features (in emission or absorption) can affect severely the intensity of the \foii~$\lambda\lambda7319+20+30+31$ lines if the sky emission is not correctly corrected. This can be especially important in low  resolution spectra. 

Photoionisation models predict a linear \tel(\foii)--\tel(\foiii) relation with a slope of between 0.7 and 0.8 \citep{Campbell:86,Garnett:92,Deharveng:00}, although some authors provide a non-linear fit between both diagnostics but with a slight curvature \citep{Pagel:92, Thuan:95,Izotov:97b,Izotov:06}. Linear fits to observational data from star-forming regions available in the literature give quite different values of the slope, although always less than or close to unity. \citet{Pilyugin:06,Pilyugin:09} give values of 0.72 and 0.84, respectively, \citet{Esteban:09} obtain 0.99, \citet{Zurita:21} 0.63, and \citet{Cataldi:25}, who uses a very extensive compilation of data from \hii\ regions and SFGs -- they do not indicate the total number of objects -- find a slope of 0.54. One striking aspect of \citet{Cataldi:25} work is that they use the collision strengths of \citet{Palay:12} for O$^{++}$ and, therefore, for the calculation of \tel(\foiii) in their objects. This is a surprising choice because several authors have indicated that those data may not be entirely correct \citep{Storey:14,JuandeDios:17,Morisset:20}. In fact, the collision strengths of \citet{Palay:12} give  \tel(\foiii) values several hundred K lower than all other available atomic data sets\footnote{\citet{Morisset:20} say, quoting \citet{Storey:14}, that the \foiii~$\lambda 4363/\lambda 5007$ line intensity ratio that is used to derive \tel(\foiii) is about 25\% higher and that it is caused by \citet{Palay:12} neglecting the 2p$^4kl$ free channels in the close-coupling expansion of the ion-electron system, that leads to an energy downshift of the broad 2p$^5$ resonance.}. Finally, \citet{Scholte:26} present the linear fit of the \tel(\foii)--\tel(\foiii) relation with the largest number of objects: 10124 SFGs, obtaining a slope of 0.648.

In the bottom panel of Fig.~\ref{fig:TN2-TO3_TO2-TO3} we show the \tel(\fnii)--\tel(\foiii) relation for 204 DESIRED-E spectra (185 \hii\ regions and 19 SFGs). The distribution of the data points shows a considerably smaller dispersion around the linear fit than the \tel(\foii)--\tel(\foiii) relation. Their ODR linear fit  is: 
\begin{equation}
\label{eq:TN2-TO3}
    T_{\rm e}(\text{\fnii}) = (0.763\pm0.032)\times T_{\rm e}(\text{\foiii}) + (2530\pm280)~{\rm K}. 
\end{equation}

The value of the Pearson coefficient of the linear fit is 0.70, which indicates a moderate-strong correlation. For \tel(\fnii), the total and intrinsic dispersions about the linear fit for this \tel--\tel\ relation are 1360 K and 790 K, respectively, significantly lower than those of \tel(\foii)--\tel(\foiii) and \tel(\fsii)--\tel(\foiii) relations, most likely due to the lower dependence of \tel(\fnii) on \nel, as discussed above and in the previous subsection. For \tel(\foii), the scatter is larger, with total and intrinsic dispersions of 1780 K and 1040 K, respectively. Photoionisation models tend to obtain that \tel(\foii) $\approx$ \tel(\fnii) \citep{Campbell:86, Garnett:92, Pagel:92, Thuan:95,Izotov:97b}. In this sense, the slope of $~$0.7 that \citet{Campbell:86} and \citet{Garnett:92} proposed for the \tel(\foii)--\tel(\foiii) relation would also be applicable for the \tel(\fnii)--\tel(\foiii) one, so our linear fit based on observations is very consistent with what is predicted by the models. However, the slope of 1.1 obtained by \citet{Scholte:26} for this \tel--\tel\ relation from data of 106 SFGs is larger than our value and that expected by photonization models.

\begin{figure}[ht!]
\centering    
\includegraphics[width=\hsize ]{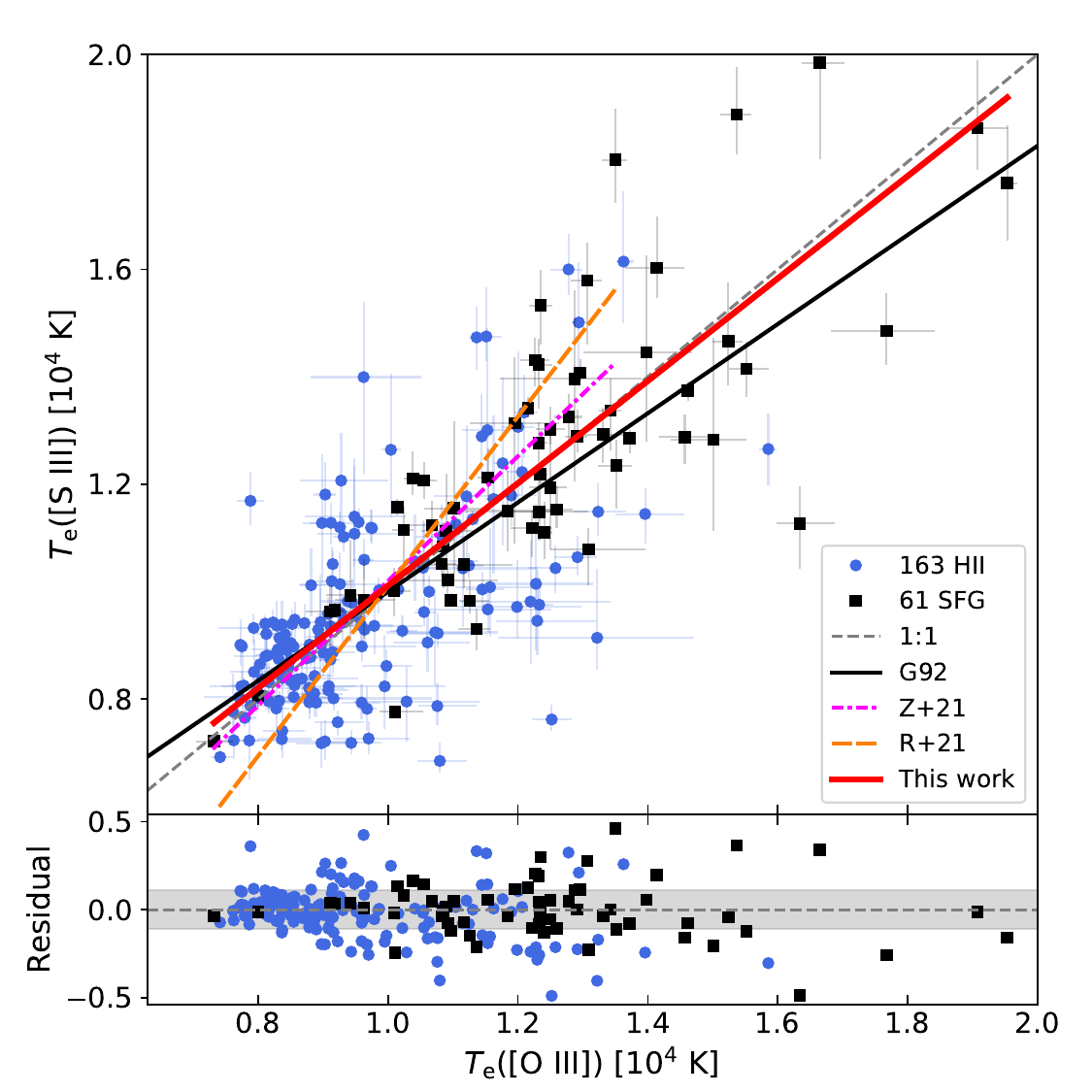}
\includegraphics[width=\hsize ]{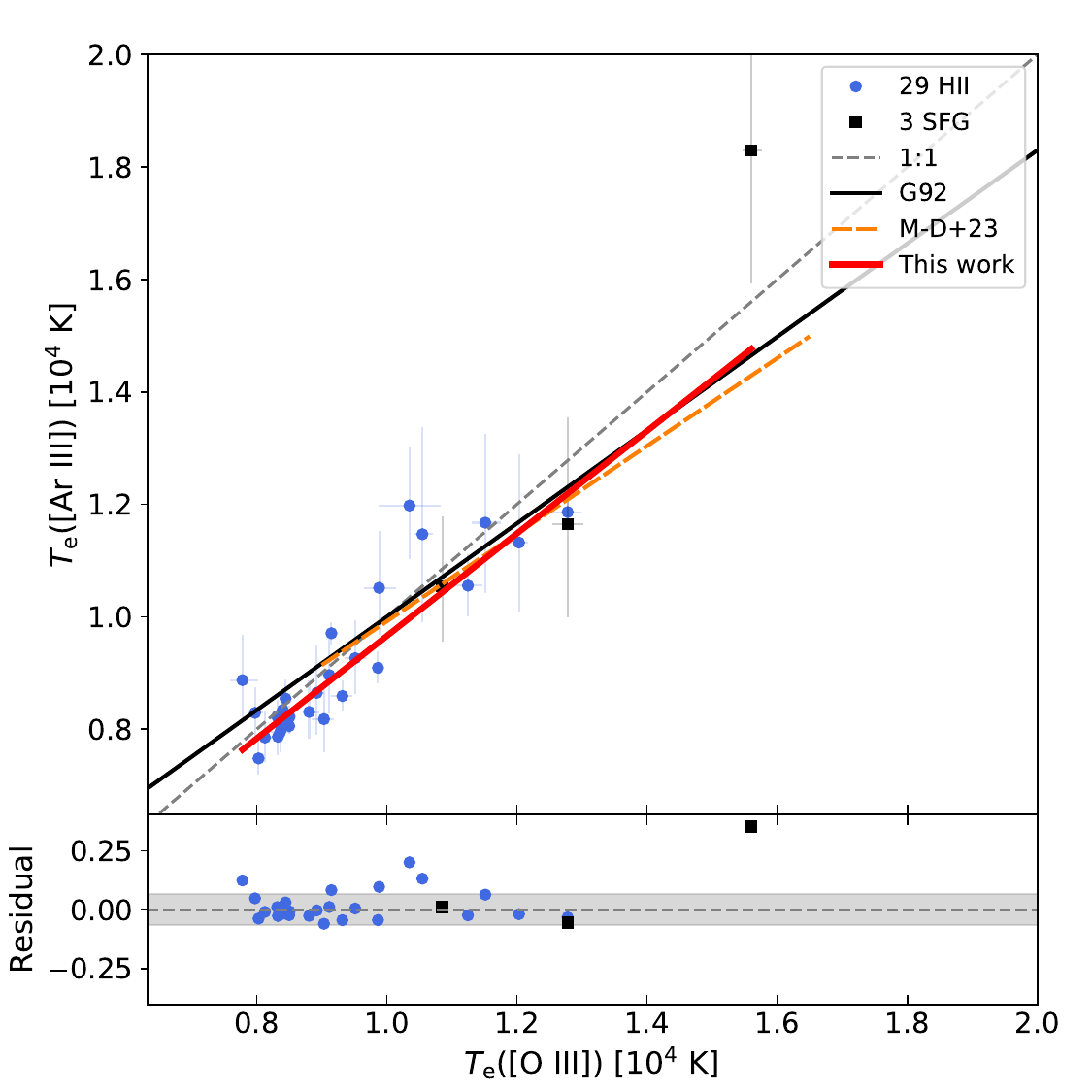}
\caption{\tel(\fsiii)--\tel(\foiii) (top) and \tel(\fariii)--\tel(\foiii) (bottom) relations obtained for our DESIRED-E sample. The red continuous lines represent the ODR linear fits to the data. The grey dashed lines show the 1:1 relation. The black continuous lines represent the linear fits obtained from photoionisation models by \citet{Garnett:92}. In the upper panel, the orange dashed line represent the linear fits to the observational data of \citet{Rogers:21}, while in the bottom panel represents de linear fit obtained from BOND models \citep{ValeAsari:16} by \citet{MendezDelgado:23b}. The rectangular inset below each plot shows the residuals to the ODR linear fit, with the grey band showing the total dispersion ($\sigma_{tot}$) of the data around the fit. } 
\label{fig:TS3-TO3_TAr3-TO3}
\end{figure}

As it is illustrated in the top panel of Fig.~\ref{fig:TS3-TO3_TAr3-TO3}, the \tel(\fsiii)--\tel(\foiii) relation defined by the DESIRED-E spectra is also fairly tight. The ODR linear fit to the 224 data points (163 \hii\ regions and 61 SFGs) is: 
\begin{equation}
\label{eq:TS3-TO3}
    T_{\rm e}(\text{\fsiii}) = (0.953\pm0.015)\times T_{\rm e}(\text{\foiii}) + (580\pm150)~{\rm K}. 
\end{equation}
\noindent The Pearson coefficient of the fit is 0.79, indicating a quite strong correlation between \tel(\fsiii) and \tel(\foiii). The total and the intrinsic dispersions of this relation for \tel(\fsiii) are quite large, 1510 K and 1280 K, respectively (1580 K and 1340 K for \tel(\foiii)). The slope of the linear fit is  fairly consistent with that expected using photoionisation codes, as it was also found in other previous studies based on observational data \citep[e.g.][]{Kennicutt:03, Bresolin:09, Zurita:12}. For example, \citet{Garnett:92} found a value of 0.83, while \citet{MendezDelgado:23b}, making use of the BOND models \citep{ValeAsari:16}, obtained a slope of 0.95. However, the consistency with models predictions -- as our results indicate -- is not commonly found in previous observation-based \tel(\fsiii)--\tel(\foiii) relations. Several works found slightly higher slopes, of the order of 1.15, as is the case of \citet{Vermeij:02}, \citet{Hagele:06} or \citet{Zurita:21}. On the other hand, \citet{Binette:12} reported a substantially higher slope, of 1.41, based on data for a sample of 102 \hii\ regions and SFGs published by \citet{PerezMontero:06} and \citet{Hagele:06}. \citet{Binette:12} find that \tel(\foiii) are systematically higher than the corresponding \tel(\fsiii) in objects with \tel(\fsiii) lower than 14,000 K. A behaviour that is not confirmed in our data. Finally, the various works of the CHAOS project that analyze the \tel(\fsiii)--\tel(\foiii) relation in \hii\ regions of different spiral galaxies find slopes between 1.27 and 1.80 and rather high intrinsic dispersions, mainly introduced by their \tel(\foiii) determinations \citep[e.g.][]{Croxall:16,Berg:20,Rogers:21}. Finally, \citet{Scholte:26}, using a large sample of 3781 SFGs, find a slope of 1.062, a value not very different from ours. 

\tel(\fariii)--\tel(\foiii) is the last \tel--\tel\ relation we discuss in this paper, which we show in the bottom panel of Fig.~\ref{fig:TS3-TO3_TAr3-TO3}. It is the one that includes the fewest observational points (32 DESIRED-E spectra, of which 29 are \hii\ regions and 3 SFGs) due to the extreme weakness of the auroral \fariii~$\lambda$5192 line. The ODR fit to these \tel\ values is: 
\begin{equation}
\label{eq:TAr3-TO3}
    T_{\rm e}(\text{\fariii}) = (0.911\pm0.069)\times T_{\rm e}(\text{\foiii}) + (540\pm610)~{\rm K}. 
\end{equation}
The relation between both \tel\ indicators is very tight, with a Pearson coefficient of 0.93, the highest of all the relations studied in this work. Furthermore, the total and intrinsic dispersions of this relation are the lowest we obtain, 890 K and 160 K, respectively, indicating that most of the dispersion comes from the observational errors. Photoionisation models predict a relation between \tel(\fariii) and \tel(\foiii) close to the 1:1 relation; \citet{Garnett:92} gives a slope of 0.83 and \citet{MendezDelgado:23b} obtain 0.95, consistent with our slope obtained from the observational data compiled in DESIRED-E. The only previous study that studied \tel--\tel\ relations involving observational determinations of \tel(\fariii) is that of \citet{MendezDelgado:23b}, which included results for 17 \hii\ regions and SFGs of the original DESIRED sample, obtaining a slope consistent with ours within the errors.

\section{The application of \tel--\tel\ relations}
\label{sec:application}

As it has been commented in Sect.~\ref{sec:introduction}, most of the works  dedicated to determining the abundances of ionised gas in star-forming regions use a two-zone (or even a three-zone in some cases) approach, but in many cases, there is only one \tel\ indicator for a single ionisation zone. Therefore, we have to apply a \tel--\tel\ relation to estimate a representative \tel\ of the ionisation zone for which we lack a measured indicator. In SFGs, with generally subsolar metallicities, the most common situation is to only have \tel(\foiii). In \hii\ regions in environments with metallicity close to solar, \tel(\foiii) is also very common as the only available \tel\ indicator, although \tel(\fnii) may also be well-measured, as is the case of Galactic \hii\ regions \citep[e.g.][]{ArellanoCordova:20b, ArellanoCordova:21} or those of the nearest spiral galaxies studied in the CHAOS project \citep[e.g.][]{Berg:15, Berg:20, Croxall:16, Rogers:21,Rogers:22}. When the available spectra include the reddest part of the optical range, between 7000 and 10,000 \AA, it is possible to determine \tel(\foii) and/or \tel(\fsiii). These indicators are easier to obtain than \tel(\fnii) because the  \foii~$\lambda\lambda7319, 7320, 7330, 7331$ and \fsiii~$\lambda6312$ auroral lines are commonly brighter than \fnii~$\lambda5755$. On the other hand, the \fsii~$\lambda\lambda$4069, 4076 auroral lines are generally rather faint -- with an intensity similar to \fnii~$\lambda5755$ -- but are in the blue part of the optical spectrum, where detectors are commonly less efficient and the line intensities more attenuated by dust extinction. Even so, the number of objects we have in DESIRED-E with measurements of \tel(\fsii) is quite high, as we indicated in the first paragraph of Sect.~\ref{subsec:low_ion}. Another situation in which we have to apply \tel--\tel\ relations is when our instrument does not cover a part of the optical spectral range, as is the case, for example, with MUSE on the VLT, where we cannot observe lines below 4650 \AA, which prevents us from obtaining \tel(\foiii), \tel(\foii) and \tel(\fsii) in local star-forming regions. Therefore, direct abundance determinations obtained with MUSE can only be based on \tel(\fsiii) and/or \tel(\fnii) for those objects \citep[e.g.][]{Groves:23,Brazzini:24,RickardsVaught:24}.

As discussed throughout Sect.~\ref{sec:Te-Te_relations}, some of the studied \tel--\tel\ relations allign more closely with photoionisation models predictions than others. In the case of relations involving \tel\ indicators of the low ionisation zone: \tel(\foii), \tel(\fnii), and \tel(\fsii), we find that 
the \tel(\fsii)--\tel(\foii) relation has a slope fairly consistent with the one predicted by photoionisation models, although with a constant offset, in the sense that \tel(\foii) is systematically on the order of 400 K larger than \tel(\fsii). However, 
the linear fits to our observational \tel(\foii)--\tel(\fnii) and \tel(\fsii)--\tel(\fnii) relations give very similar slopes and deviate in a similar way from the relation predicted by photoionisation models. The general trend is that \tel(\foii) and \tel(\fsii) tend to be larger than \tel(\fnii) as \tel(\fnii) increases, although it shows a change in behaviour at \tel(\fnii) > 12,000 K, which can be clearly seen  in the residual insets of the \tel--\tel\ relations shown in Fig.~\ref{fig:TO2-TN2_TS2-TN2}. In these insets, the residuals of the ODR linear fits to the \tel(\foii)--\tel(\fnii) and \tel(\fsii)--\tel(\fnii) distributions are systematically negative for \tel(\fnii) > 12,000 K. 

However, the number of objects with available \tel(\fnii) determinations at such high \tel\ values is very limited and could be affected by observational bias. In fact, objects with higher \tel\ tend to have lower metallicity and a higher ionisation degree, so the amount of N$^+$ is very small and, therefore, their \fnii~$\lambda$5755/H$\beta$ intensity ratio very small. This means that the value of \tel(\fnii) of our observational points tends to be  correlated with its error. Moreover, it is even possible that the intensity of \fnii~$\lambda$5755 may have been overestimated in some high-\tel\ objects \citep[e.g.][]{Rola:94}. 
On the other hand, inspecting Table~\ref{table:ODR_fits}, we can see how the \tel--\tel\ relations involving \tel(\fnii) consistently show $\sigma_{int}$ values on the order of a factor of two less to those involving \tel(\foii) or \tel(\fsii). Considering the discussion outlined in Sect.~\ref{subsec:low_ion}, the results described in this paragraph can be explained by the different dependence of \tel(\foii), \tel(\fsii), and \tel(\fnii) on \nel\ and the presence of \nel\  inhomogeneities in the ionised gas. The possible effects of the degree of ionization are discussed briefly in Appendix~\ref{sec:appendix_c}.

As a first recommendation, we propose that when \tel(\foiii) and/or \tel(\fsiii) are available, but not a low-ionisation zone indicator, the best option is to use the \tel(\fnii)--\tel(\foiii) and/or \tel(\fnii)--\tel(\fsiii) relations to obtain \tel(\fnii). This choice guarantees an assumed \tel\ value with less intrinsic error for the low-ionisation zone. Photoionisation models agree that \tel(\foii) $\approx$ \tel(\fnii) \citep{Campbell:86, Garnett:92, Thuan:95, Izotov:97b}, so using \tel(\fnii) for calculating the abundance of low-ionisation ions, especially the O$^+$/H$^+$ ratio, seems a more than reasonable approximation. In any case, it should be noted that, because the number of determinations is very small at high \tel(\fnii) values, the \tel(\fnii)--\tel(\fsiii) relation can be considered valid for \tel(\fsiii) $<$ 16,000 K and the \tel(\fnii)--\tel(\foiii) relation for \tel(\foiii) $<$ 14,000 K; above those values the relations should be considered quite unreliable.
As discussed in Sect.~\ref{subsec:rel_S3} and \ref{subsec:rel_O3}, our linear ODR fits to the \tel(\fnii)--\tel(\fsiii), \tel(\foii)--\tel(\fsiii), \tel(\fnii)--\tel(\foiii), and \tel(\fariii)--\tel(\foiii) distributions are fairly consistent with the predictions of photoionisation models \citep[e.g.][]{Garnett:92, MendezDelgado:23b}, while the fits to the \tel(\foii)--\tel(\foiii) and \tel(\fsiii)--\tel(\foiii) ones do not show such a good consistency, but a reasonable one in any case. It is worth noting that the degree of consistency with the photoionisation models tends to increase with decreasing values of the $\sigma_{int}$ of the ODR linear fit. 

\section{Conclusions}
\label{sec:conclusions}

In this work we have carried out a homogeneous analysis of \tel\ derived from six optical CEL diagnostics -- \tel(\fnii), \tel(\foii), \tel(\fsii), \tel(\fsiii), \tel(\foiii), and \tel(\fariii) -- using 699 spectra of Galactic and extragalactic {\hii} regions and star-forming galaxies compiled in the DESIRED-E database. We recomputed \nel\ and \tel\ with updated atomic data and a consistent methodology, allowing us to explore the behaviour of the different \tel--\tel\ relations. Our main results can be summarised as follows:

\begin{itemize}

\item [1.] Relations involving low-ionisation diagnostics exhibit significant large $\sigma_{int}$ values. In particular, the relations involving the \tel(\foii) and \tel(\fsii) diagnostics show the largest dispersions, which can be explained by their stronger sensitivity to \nel\ inhomogeneities, possible recombination contributions, and observational uncertainties related to the large wavelength baselines involved on these diagnostics.

\item [2.] \tel--\tel\ relations with   \tel(\fnii) in the y-axis give  $\sigma_{int}$ values consistently lower than the equivalent relations involving   \tel(\foii) or \tel(\fsii). This may be due to the lower \nel-sensitivity of this \tel\ diagnostic. We recommend using relations involving \tel(\fnii) instead of  \tel(\foii) or \tel(\fsii) to estimate a representative \tel\ of the low-ionisation zone when only \tel(\foiii) and/or \tel(\fsiii) are available, due to the apparent greater reliability of \tel(\fnii). Unfortunately, \tel(\fnii) can be extremely faint and very difficult to detect at the low metallicities typical of SFGs. 

\item [3.] Among the relations involving intermediate- and high-ionisation diagnostics, 
\tel(\fariii)--\tel(\foiii) shows the tightest relation obtained in this study, although the number of available measurements of the auroral {\fariii} $\lambda5192$ line is very limited.

\item [4.] The slopes of the ODR linear fits applied to the \tel--\tel\ relations are broadly consistent with those predicted by photoionisation models, especially in the cases of \tel(\fnii)--\tel(\fsiii), \tel(\foii)--\tel(\fsiii), \tel(\fnii)--\tel(\foiii), and \tel(\fariii)--\tel(\foiii) relations. The degree of agreement improves for relations with lower $\sigma_{int}$, suggesting that physical and observational limitations dominate the discrepancies.

\item [5.] The set of empirical \tel--\tel\ relations presented here provides a homogeneous observational foundation for estimating \tel\ in objects where only one diagnostic is available. These relations are particularly valuable for the determination of chemical abundances in local extragalactic {\hii} regions -- our sample is significantly much more limited for SFGs and, in principle, less applicable for these objects --, where \tel(\foiii) is often the only \tel\ diagnostic available.

\end{itemize}

Overall, this study provides the most comprehensive observational characterisation to date of the relationships among the different \tel\ diagnostics in local \hii\ regions. As the DESIRED-E database continues to expand, future work combining these data with deep spatially resolved spectroscopy of star-forming regions will enable an increasingly refined view of the \tel\ structure in ionised regions of the local Universe.

\section*{Data Availability}

The complete tables of references, physical
conditions and chemical abundances for all the objects in our sample can be found at: \href{https://zenodo.org/records/19471717}{https://zenodo.org/records/19471717}.

\begin{acknowledgements}
MOG, CE, JGR and ERR acknowledge financial support from the Agencia Estatal de Investigaci\'on of the Ministerio de Ciencia e Innovaci\'on  y Universidades (AEI-MCIU) under grant ``The internal structure of ionised nebulae and its effects in the determination of the chemical composition of the interstellar medium and the Universe'' with reference PID2023-151648NB-I00 (DOI:10.13039/5011000110339). JGR also acknowledges support from the AEI-MCIU and from the European Regional Development Fund (ERDF) under grant ``Planetary nebulae as the key to understanding binary stellar evolution'' with reference PID2022-136653NA-I00 (DOI:10.13039/501100011033). AZLA gratefully acknowledges the support provided by the Postdoctoral Program (POSDOC) of UNAM (Universidad Nacional Autónoma de México). JEMD and AZLA also acknowledge support from the SECIHTI project CBF-2025-I-2048, ``Resolviendo la Física Interna de las Galaxias: De las Escalas Locales a la Estructura Global con el SDSS-V Local Volume Mapper''. JEMD, MOG, CE, JGR, AZLA, KZAC, FFRO and ERR thank the support by UNAM/DGAPA/PAPIIT/IA103326 project ``DESIRED (DEep Spectra of ionised Regions Database): de las emisiones más sutiles a la física fundamental del universo’’ .\end{acknowledgements}

\bibliographystyle{aa}
\bibliography{refs} 


\begin{appendix} 

\section{References for the spectroscopic data}
\label{sec:appendix_b}

The complete reference tables associated with Appendix A can be found at: \href{https://zenodo.org/records/19471717}{https://zenodo.org/records/19471717}

The references for the spectroscopic data are: \citet{ArellanoCordova:21}; \citet{Berg:13}; \citet{Bresolin:07a}; \citet{DelgadoInglada:16}; \citet{DominguezGuzman:22}; \citet{Esteban:04, Esteban:09, Esteban:13, Esteban:14, Esteban:17, Esteban:20}; \citet{Esteban:18}; \citet{Fernandez:18, Fernandez:22}; \citet{FernandezMartin:17}; \citet{GarciaRojas:04, GarciaRojas:05, GarciaRojas:06, GarciaRojas:07}; \citet{Guseva:00, Guseva:03c, Guseva:09, Guseva:11, Guseva:24}; \citet{Izotov:94, Izotov:97b, Izotov:04b, Izotov:06, Izotov:09, Izotov:17b, Izotov:21b}; \citet{Izotov:04}; \citet{Kurt:99}; \citet{LopezSanchez:07, Lopez-Sanchez:09}; \citet{MendezDelgado:21a, MendezDelgado:21b, MendezDelgado:22b}; \citet{MesaDelgado:09}; \citet{Noeske:00}; \citet{Peimbert:86, Peimbert:05, Peimbert:12}; \citet{Peimbert:03}; \citet{PenaGuerrero:12}; \citet{Rogers:22}; \citet{Skillman:03}; \citet{Thuan:95, Thuan:05}; \citet{Toribio:16}; \citet{Valerdi:19, Valerdi:21}; \citet{Zurita:12}

\section{Median electron temperatures for each ion and group of objects}
\label{sec:appendix_a}
\begin{figure*}[ht!]
\centering    
\includegraphics[scale=0.40]{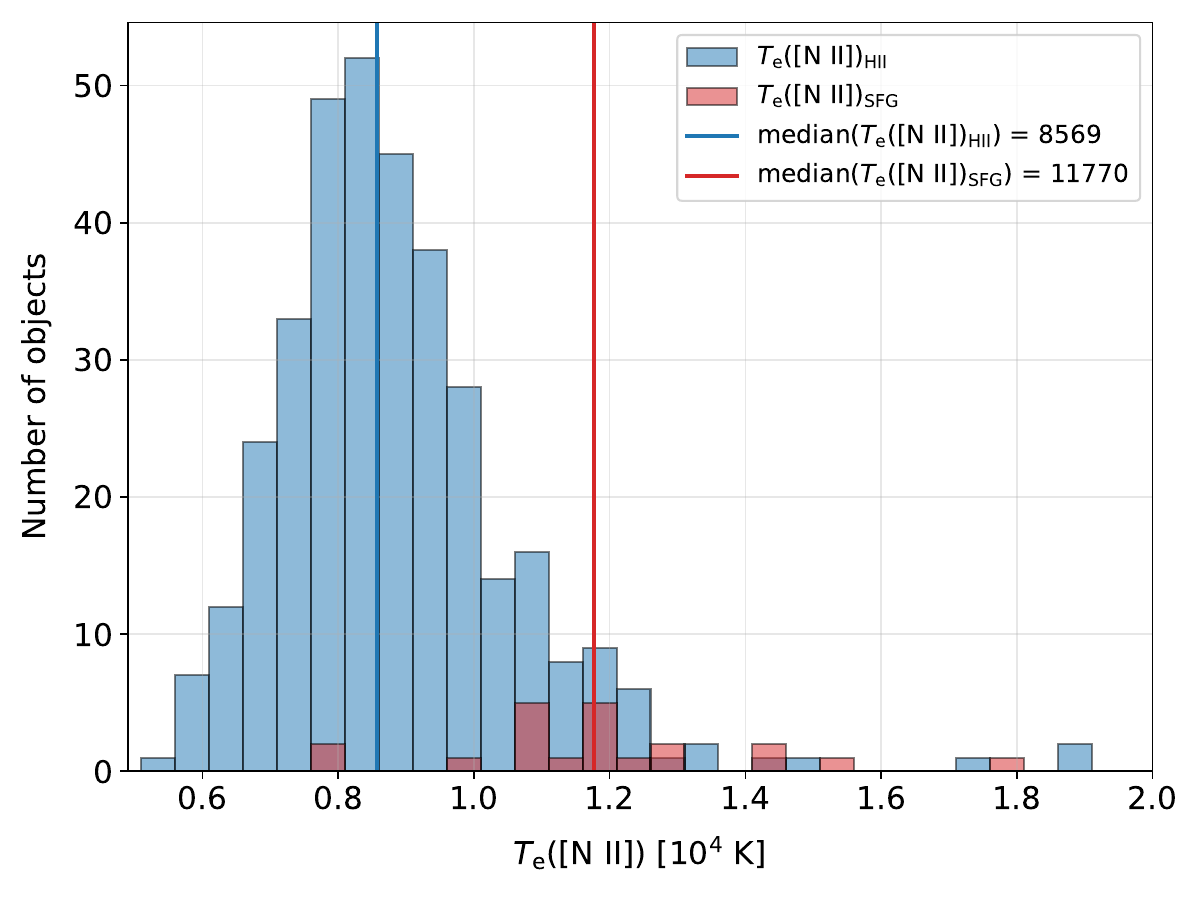}
\includegraphics[scale=0.40]{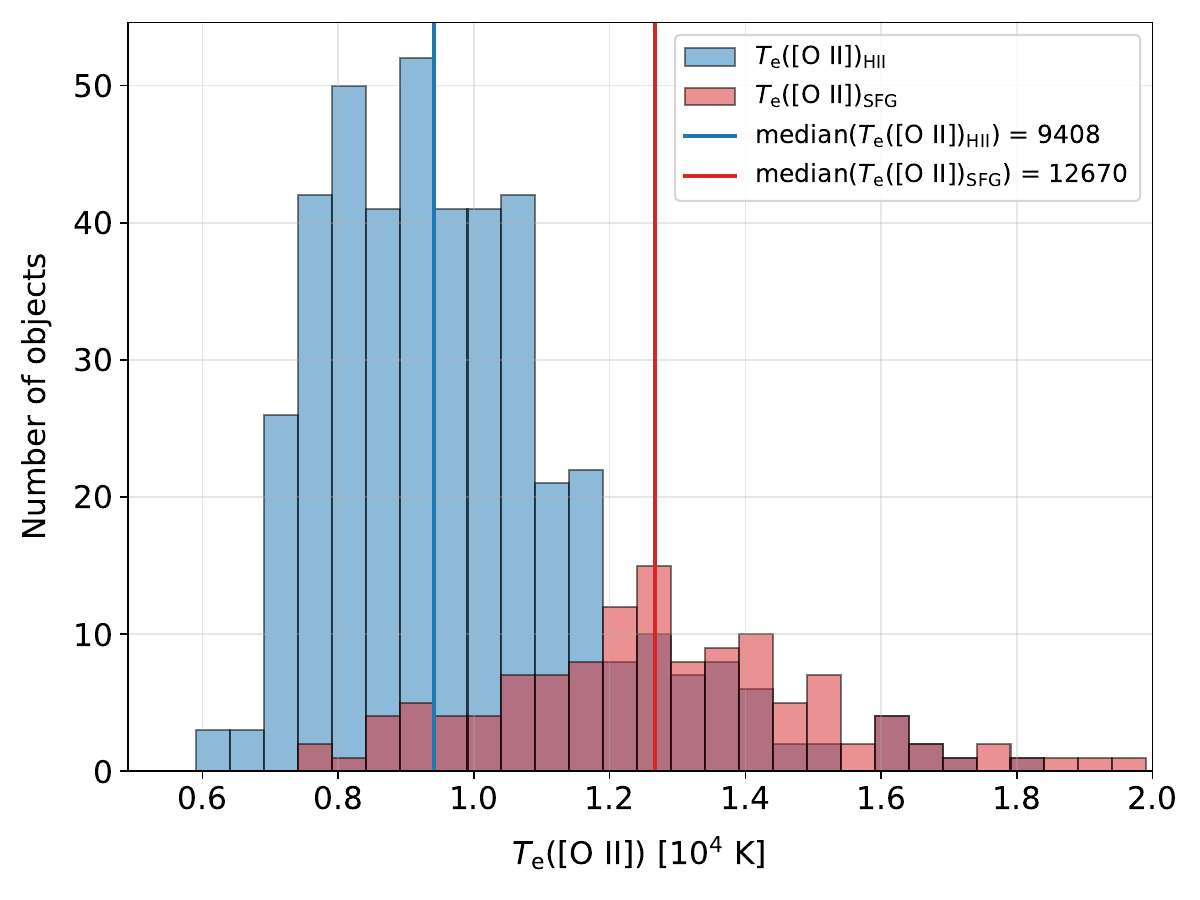}

\includegraphics[scale=0.40]{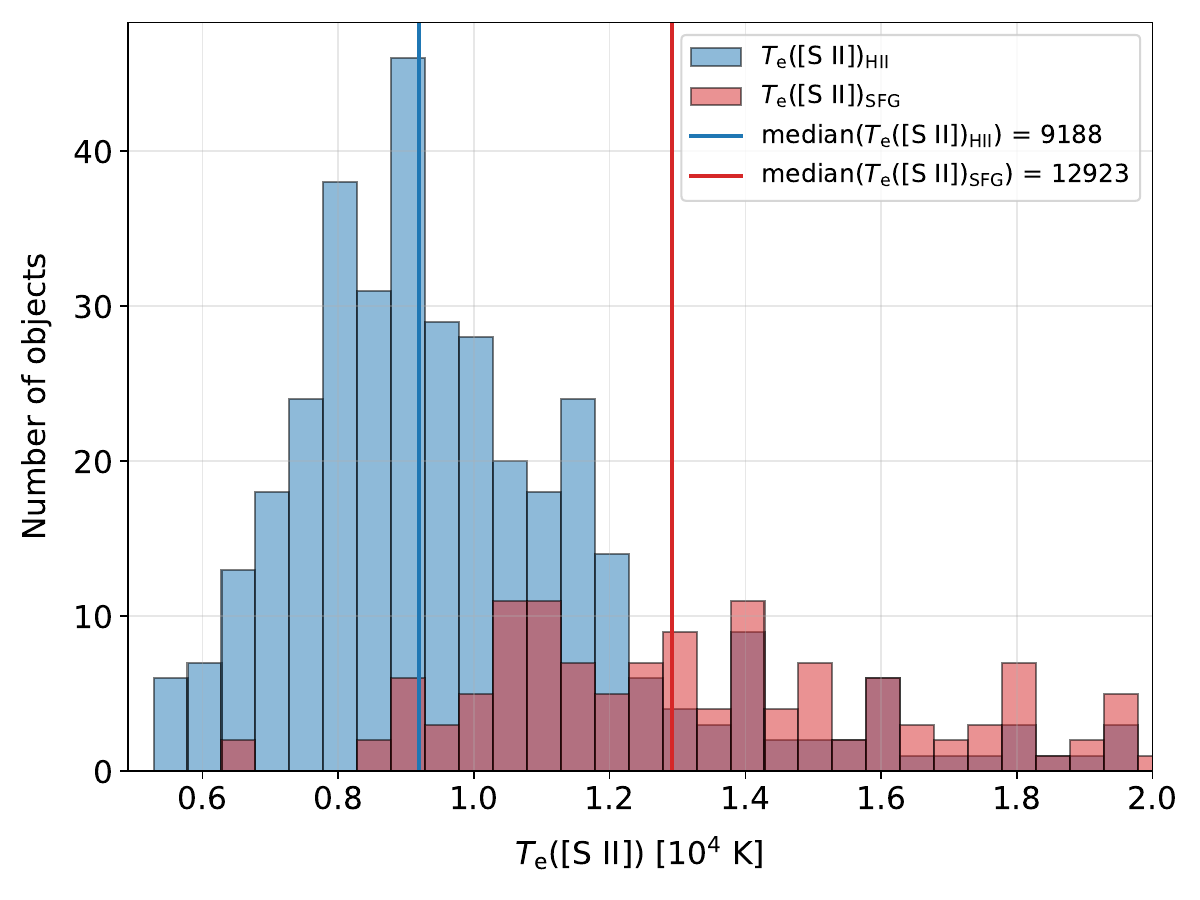}
\includegraphics[scale=0.40]{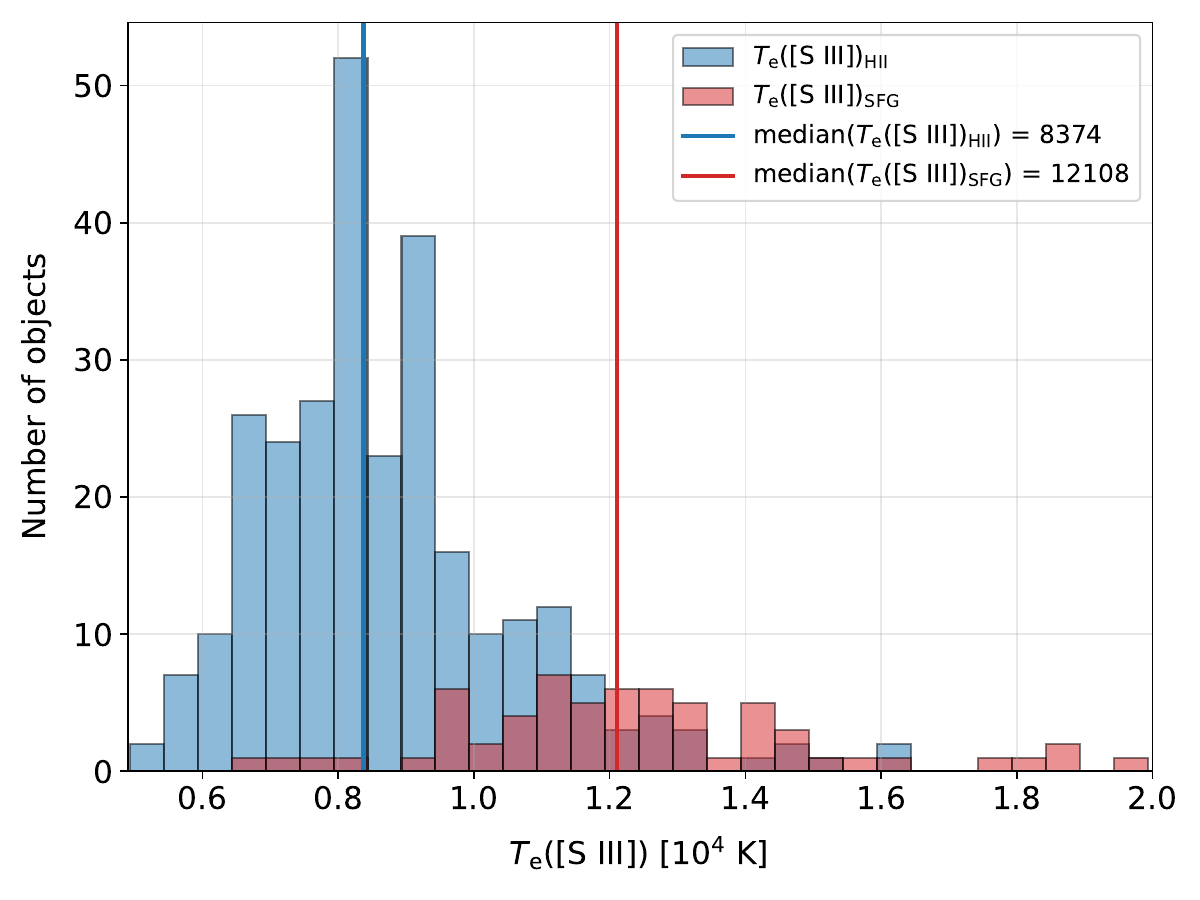}

\includegraphics[scale=0.40]{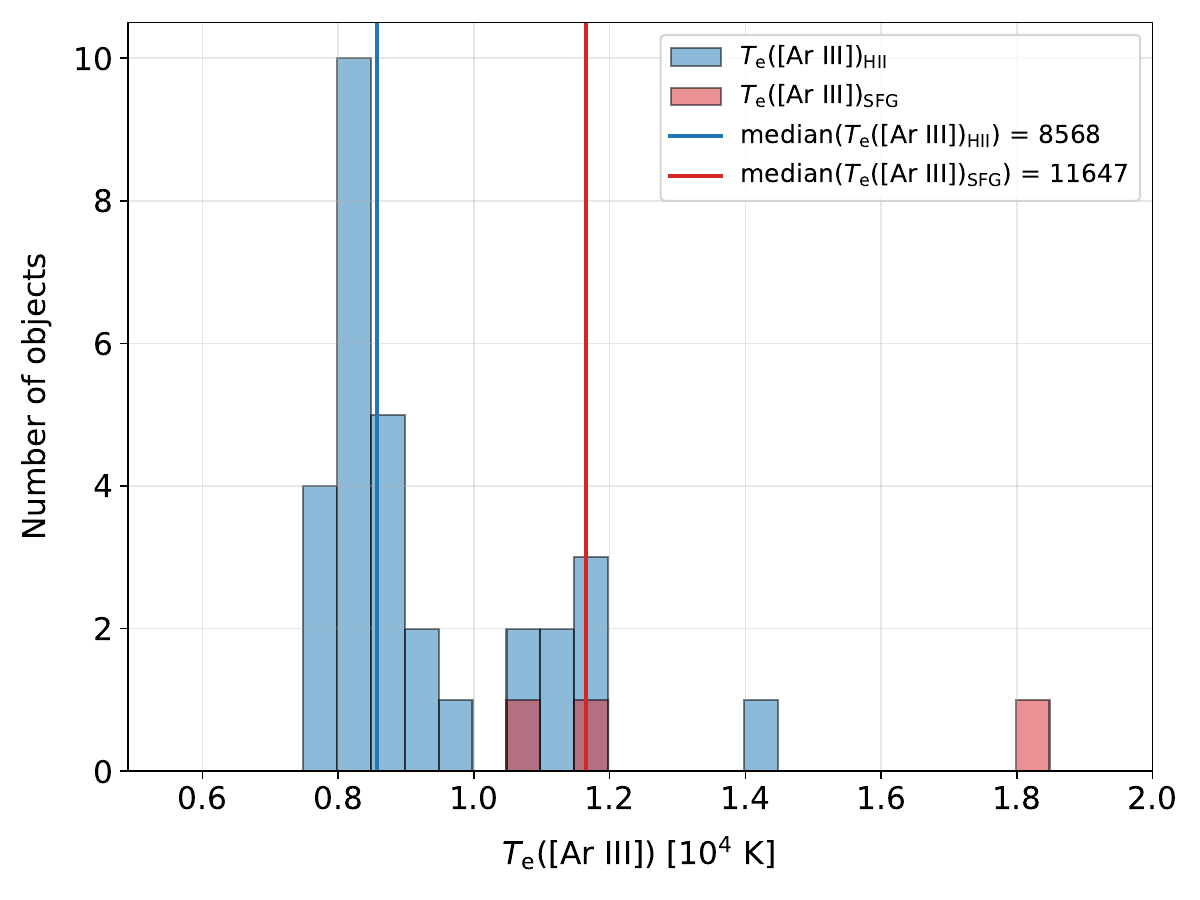}
\includegraphics[scale=0.40]{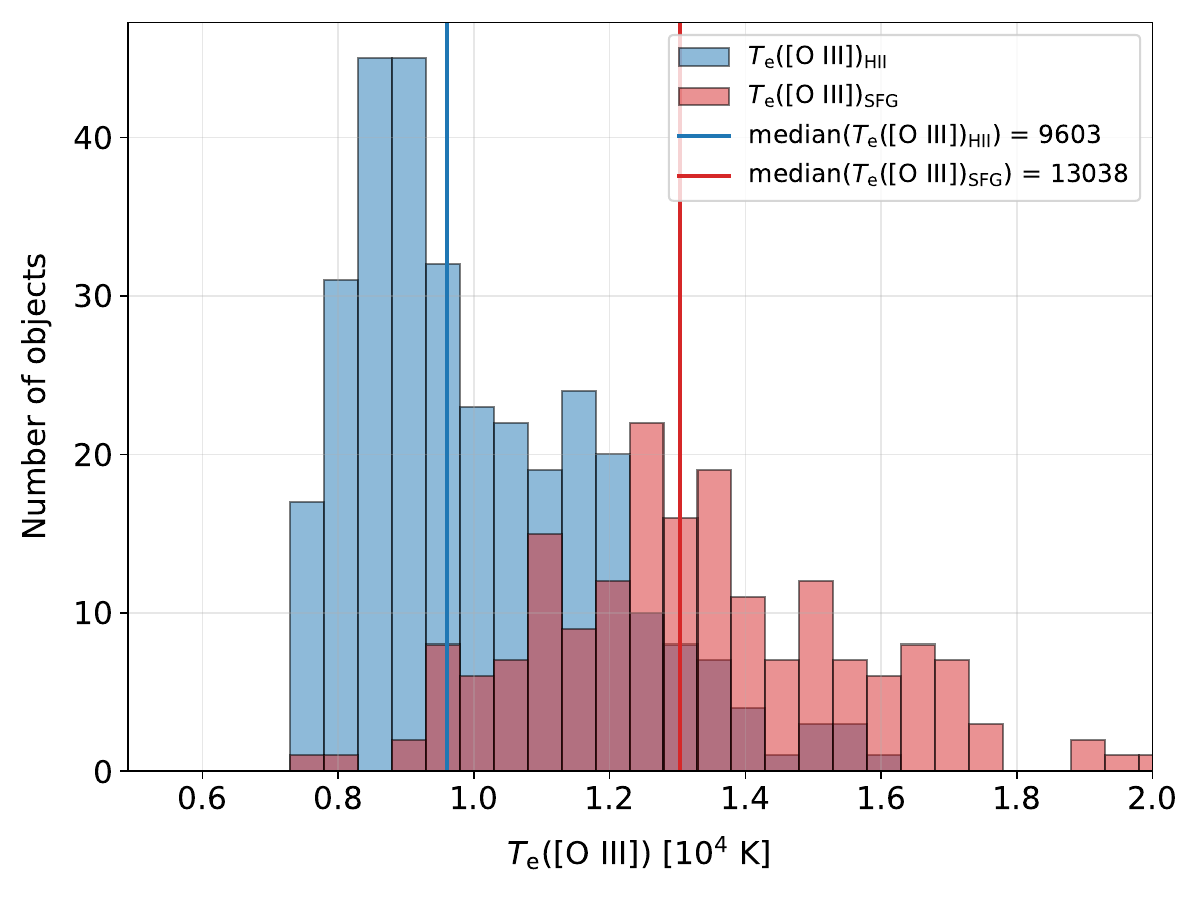}
\caption{Histograms of the different \tel\ indicators used in this paper separated by group of objects: \hii\ regions (blue bars) and SFGs (red bars). We use 500 K wide bins. The blue and red vertical lines show the median \tel\ values for \hii\ regions and SFGs, respectively. The number of objects represented in each diagram and the median values of \tel\ and 12+log(O/H) are given in Table~\ref{table:Median_Tes}} 
\label{fig:Median_Tes}
\end{figure*}

In Fig.~\ref{fig:Median_Tes}, we show the histograms of the different \tel\ indicators used in this paper, \tel(\fnii), \tel(\foii), \tel(\fsii), \tel(\fsiii), \tel(\foiii), and \tel(\fariii) separated by group of objects: \hii\ regions and SFGs. The median of the different \tel\ values separated by group of objects, the number of objects represented in each case, and their median value of 12+log(O/H), are included in Table~\ref{table:Median_Tes}. From Fig.~\ref{fig:Median_Tes} and Table~\ref{table:Median_Tes} it is clear that \hii\ regions and SFGs show a quite different median value of any \tel\ considered, being always larger in the case of SFGs. This is mainly due to their lower median metallicity. While the representative median value of 12+log(O/H) of our sample of \hii\ regions is around 8.41, the median drops to 8.06 in the case of SFGs. It is  striking that, regardless of the number of objects -- which can be very different for the different \tel\ indicators -- the ratio between the median \tel\ values of SFGs and \hii\ regions (last column of Table~\ref{table:Median_Tes}) is significantly constant, between 1.35 and 1.45, regardless of the atom and ionisation state we consider. This indicates that all the \tel\ indicators exhibit a basically similar dependence on metallicity. 

\begin{table*}
    \centering
    \caption{Median of the different \tel\ indicators and group of objects}
    \label{table:Median_Tes}
    \begin{tabular}{ccccccccc}
        \hline
        \noalign{\smallskip}
        & \multicolumn{3}{c}{\hii\ regions} & & \multicolumn{3}{c}{SFGs} & \\
        \cline{2-4} \cline{6-8}
        & Median & & Median & & Median & & Median & \tel\ ratio \\
        \tel & \tel & No. & 12+log(O/H) & & \tel & No. & 12+log(O/H) & SFGs/\hii \\
        \noalign{\smallskip}
        \hline
        \noalign{\smallskip}
        \tel(\fnii) & 8569 & 350 & 8.45 & & 11770 & 21 & 8.07 & 1.37 \\
        \tel(\foii) & 9408 & 435 & 8.41 & & 12670 & 123 & 8.04 & 1.35 \\
        \tel(\fsii) & 9188 & 361 & 8.41 & & 12923 & 126 & 8.05 & 1.41 \\
        \tel(\fsiii) & 8374 & 282 & 8.43 & & 12108 & 63 & 8.09 & 1.45 \\       
        \tel(\fariii) & 8568 & 29 & 8.45 & & 11647 & 3 & 8.07 & 1.36 \\
        \tel(\foiii) & 9603 & 315 & 8.32 & & 13038 & 183 & 8.05 & 1.36 \\ 
        \noalign{\smallskip}
        \hline
    \end{tabular}
\end{table*}

Other studies that compare median values of different \tel\ indicators in a similar way are those by \citet{RickardsVaught:24} and \citet{Scholte:26}. \citet{RickardsVaught:24} present median values for \tel(\fnii), \tel(\foii), \tel(\fsii), and \tel(\fsiii) for a similar, though slightly smaller, number of \hii\ regions than ours, but significantly smaller (only 26) in the case of \tel(\foiii). The median values of \citet{RickardsVaught:24} are very similar to ours for \hii\ regions, with differences on the order of or less than 500 K for \tel(\fnii), \tel(\foii), \tel(\fsii), and \tel(\fsiii). In the case of \tel(\foiii), their median value is approximately 2400 K higher than ours, deviating much further from the values of the other \tel\ indicators. \citet{RickardsVaught:24} ordering of the median values from lowest to highest is: \tel(\fnii), \tel(\fsiii), \tel(\fsii), \tel(\foii), and \tel(\foiii), quite similar to the one we found, except that the order of the two lowest values is swapped: \tel(\fsiii), \tel(\fnii), \tel(\fsii), \tel(\foii), and \tel(\foiii) in our determinations. \citet{Scholte:26} also obtain median values for \tel(\fnii), \tel(\foii), \tel(\fsii), \tel(\fsiii), and \tel(\foiii), but for a much larger number of SFGs than we do, especially for \tel(\foii), \tel(\fsiii), and \tel(\foiii), so their results, in principle, should be statistically more significant for these objects and \tel\ indicators. Most median \tel\ values we obtain for our more limited  sample of SFGs are quite different from those of \citet{Scholte:26}. While our median of \tel(\fsiii) is identical, the ones of \tel(\fnii) and \tel(\foiii) are around 1000 K lower than the medians obtained by \citet{Scholte:26}. In the case of \tel(\foii) and \tel(\fsii), our medians are around 2500 K higher. This large difference is difficult to explain and cannot be attributed to the atomic data used. In the case of O$^+$, they are exactly the same, while for S$^+$ they differ in the transition probability reference -- \citet{Scholte:26} use \citet{Rynkun:19} --, although the use of one or the other can not account for the large difference between the median value of \tel(\fsii) reported  in both works\footnote{The use of the transition probabilities of \citet{Rynkun:19} provide values about 200 K lower than those of \citet{Irimia:05} when \tel(\fsii)$\sim$10,000 K. The difference increases at higher temperatures, being about 800 K lower when \tel(\fsii)$\sim$18,000 K}. The difference also does not appear to be necessarily related to the very different number of objects used to obtain the values for each indicator. For example, the contrast is not so pronounced in the case of \tel(\fsii). While we counted \tel(\fsii) determinations for 126 SFGs, the sample for \citet{Scholte:26} is 421, but the difference between the median values is about 2400 K, which indicates that there should be a significant systematic effect on \tel(\fsii) determination between the two studies. There is another important disparity between our set of median \tel\ values for SFGs and those reported by \citet{Scholte:26}. In our case, the maximum discrepancy among the different temperature indicators does not exceed 1300~K, whereas in the calculations of \citet{Scholte:26} the corresponding differences reach approximately 3700~K. All temperature indicators associated with the low-ionisation zone in  \citet{Scholte:26} data, namely \tel(\fnii), \tel(\foii), and \tel(\fsii), yield median values close to 10,500~K, while \tel(\foiii) shows a significantly higher median temperature of about 14,000~K. These large disparities are difficult to understand, mainly because they do not appear to be due to a simple metallicity effect. Certainly, no major differences in metallicity would be expected between the SFG subsamples used by \citet{Scholte:26} to obtain their values for each \tel\ indicator.

\section{Residuals as a Function of the ionisation Degree}
\label{sec:appendix_c}

\begin{figure*}[ht!]
\centering    

\begin{minipage}{0.3\textwidth}
    \centering
    \includegraphics[width=\linewidth]{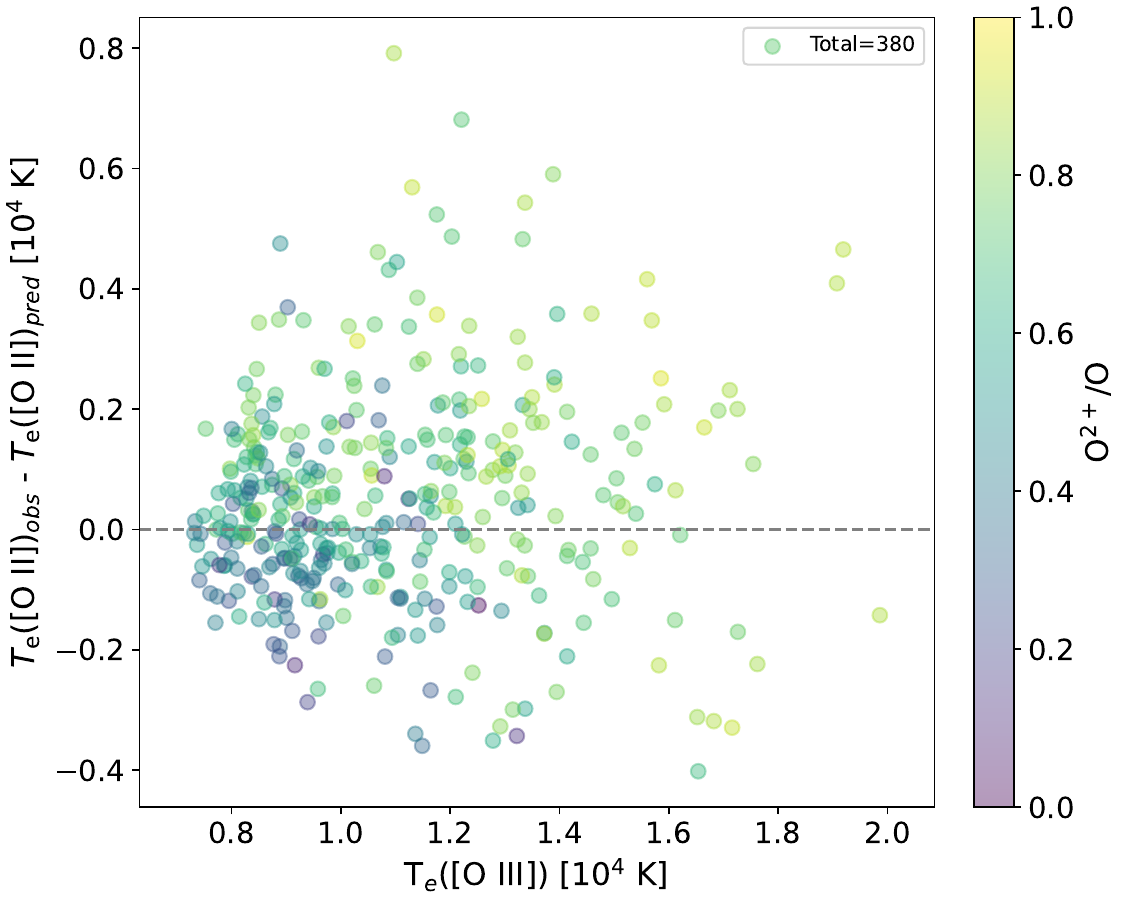}
    \label{fig:resTO2TO3}
\end{minipage}
\begin{minipage}{0.3\textwidth}
    \centering
    \includegraphics[width=\linewidth]{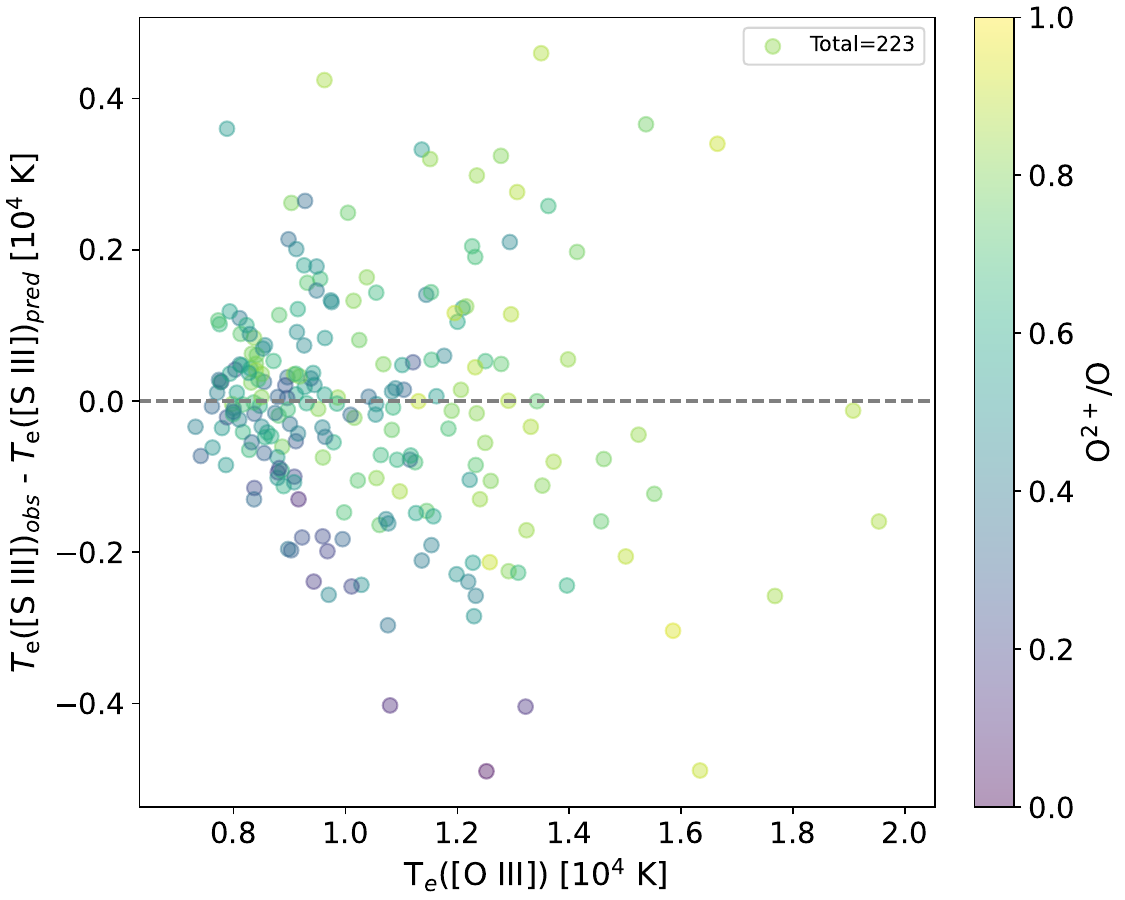}
    \label{fig:resTS3TO3}
\end{minipage}
\begin{minipage}{0.3\textwidth}
    \centering
    \includegraphics[width=\linewidth]{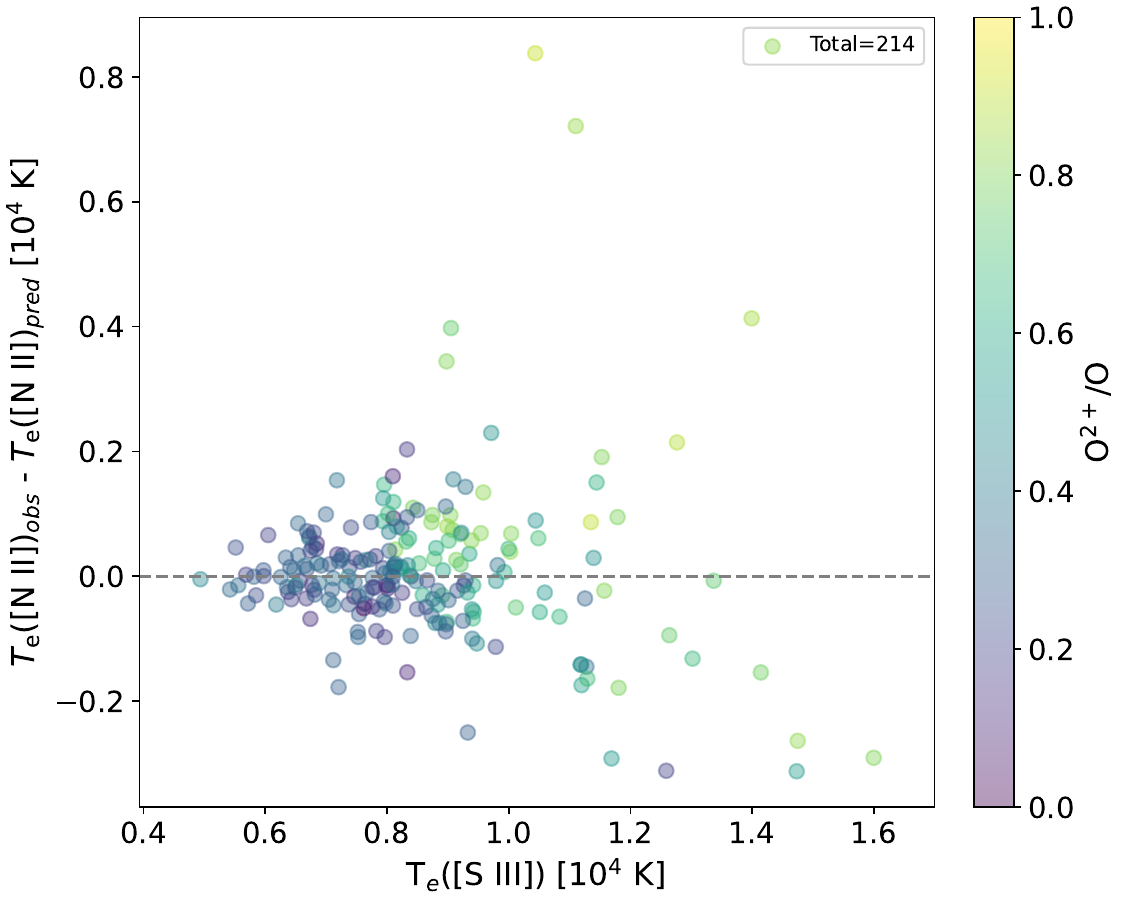}
    \label{fig:TN2TS3}
\end{minipage}

\caption{Residuals of the temperature relations as a function of the diagnostic in which the relation is based: \tel(\foii)-\tel(\foiii) (left), \tel(\fsiii)-\tel(\foiii) (center) and \tel(\fnii)-\tel(\fsiii) (right). The residuals are defined as the difference between the observed temperature and the one predicted by the corresponding relation. The color of the points varies according to their ionisation degree.}
\label{fig:Residuals}
\end{figure*}

In this Appendix we explore the possible impact of the ionisation conditions on the temperature relations analysed throughout this work. Fig.~\ref{fig:Residuals} shows the residuals, derived as $T_y-(m\times T_x+n)$, as a function of the temperature on which the relation is based.
The relations shown are \tel(\foii) and \tel(\fsiii) as functions of \tel(\foiii) (left and central panels of Fig.~\ref{fig:Residuals}, respectively), and \tel(\fnii) as a function of \tel(\fsiii) (right panel of Fig.~\ref{fig:Residuals}). The data points are color-coded according to their ionisation degree.

The ionisation degree  is used here as a proxy for the ionisation parameter, allowing us to assess whether variations in the radiation field may introduce systematic effects in the derived temperature relations. If the ionisation parameter were a dominant factor, one would expect to observe a trend in the residuals as a function of the color scale.
The diagrams show that the dispersion of the residuals increases towards higher electron temperatures. This behaviour is consistent with the fact that higher temperatures are typically associated with regions of higher excitation and, therefore, higher ionisation degree. However, despite this increase in scatter, no clear systematic dependence of the residuals on the ionisation degree is observed at fixed temperature.
In particular, the residuals do not display any monotonic trend with the ionisation parameter, nor do they show a segregation that would indicate a secondary dependence of the temperature relations on this quantity. The distribution of points remains broadly symmetric around zero across the full range of ionisation conditions.
These results indicate that, although the ionisation degree is somehow correlated with temperature and contributes to the overall increase in dispersion at the high-temperature end, it does not appear to be the primary driver of the deviations from the fitted relations.

\end{appendix}
\end{document}